\documentclass{agujournal2019}
\usepackage{url} 
\usepackage{lineno}
\usepackage{soul}
\usepackage[version=4]{mhchem}
\usepackage{lipsum}  
\usepackage{subcaption}
\usepackage{xcolor}
\usepackage{amsmath}
\usepackage{amssymb}
\usepackage{siunitx}
\usepackage{graphicx}
\usepackage{mathabx}

\DeclareSIUnit\bar{bar}
\DeclareSIUnit\AU{AU}
\DeclareSIUnit\dex{dex}
\DeclareSIUnit\erg{erg}
\DeclareSIUnit\day{day}
\DeclareSIUnit\year{yr}
\newcommand{\WPMS}{\watt\per\meter\squared}

\journalname{JGR: Planets}

\begin{document}

\title{Magma ocean evolution at arbitrary redox state}

%
%

\authors{Harrison Nicholls\affil{1}, Tim Lichtenberg\affil{2}, Dan J. Bower\affil{3,4}, and Raymond Pierrehumbert\affil{1}}

\affiliation{1}{Atmospheric, Oceanic and Planetary Physics, University of Oxford, UK}
\affiliation{2}{Kapteyn Astronomical Institute, University of Groningen, The Netherlands}
\affiliation{3}{Center for Space and Habitability, University of Bern, Switzerland}
\affiliation{4}{Institute of Geochemistry and Petrology, ETH Zurich, Switzerland}

\correspondingauthor{Harrison Nicholls}{harrison.nicholls@physics.ox.ac.uk}



\begin{keypoints}

\item Atmospheres overlying magma oceans can have diverse volatile compositions depending on mantle redox and volatile endowment.

\item Magma oceans on Earth-like planets may not solidify, retaining substantial melt fractions, and hence volatiles, in their interiors.

\item While initially convective, modelling indicates that these atmospheres may become convectively stable in the later stages of their evolution.

\end{keypoints}

%
%

\begin{abstract}
Interactions between magma oceans and overlying atmospheres on young rocky planets leads to an evolving feedback of outgassing, greenhouse forcing, and mantle melt fraction. Previous studies have predominantly focused on the solidification of oxidized Earth-similar planets, but the diversity in mean density and irradiation observed in the low-mass exoplanet census motivate exploration of strongly varying geochemical scenarios.
We aim to explore how variable redox properties alter the duration of magma ocean solidification, the equilibrium thermodynamic state, melt fraction of the mantle, and atmospheric composition.
We develop a 1D coupled interior-atmosphere model that can simulate the time-evolution of lava planets. This is applied across a grid of fixed redox states, orbital separations, hydrogen endowments, and C/H ratios around a Sun-like star.
The composition of these atmospheres is highly variable before and during solidification. The evolutionary path of an Earth-like planet at 1 AU ranges between permanent magma ocean states and solidification within 1 Myr. Recently solidified planets typically host \ce{H2O}- or \ce{H2}-dominated atmospheres in the absence of escape. 
Orbital separation is the primary factor determining magma ocean evolution, followed by the total hydrogen endowment, mantle oxygen fugacity, and finally the planet's C/H ratio. Collisional absorption by \ce{H2} induces a greenhouse effect which can prevent or stall magma ocean solidification. Through this effect, as well as the outgassing of other volatiles, geochemical properties exert significant control over the fate of magma oceans on rocky planets. 

\end{abstract}

\section*{Plain Language Summary}
The mantle of modern Earth is solid. However, young planets could have (partially) molten mantles which enable rapid transport of gases and energy to the surface. Such mantles are termed ``magma oceans''.
We explore how the properties of a planet determine magma ocean duration and study what kinds of atmospheres are produced by outgassing.
We do this by developing a computational model to simulate their evolution over time, which is then applied to an Earth-like planet over a range of different conditions.
Our simulations show that the composition of atmospheres on magma ocean planets is highly variable. However, recently solidified planets typically host hot atmospheres containing large amounts of water vapour. Evolutionary timescales also cover a wide range, with some planets maintaining permanent magma oceans while others rapidly solidify. 
While the main factor determining magma ocean evolution is orbital separation, insulation by the greenhouse effects of \ce{H2} and \ce{H2O} can act to prevent or slow down their solidification. It is important to consider the geochemical properties of planetary magma oceans, as these influence the composition of overlying atmospheres.


\section{Introduction}
\label{sec:introduction}

It is thought that the Earth, the Moon, Mars, and Venus have all been partially or entirely molten at least once in their lifetimes, maintaining a so-called ``magma ocean'' \cite{tonks_magma_1993, warren_magma_1985, elkins_magma_2008, nakajima_magma_2021}. Direct measurements of KREEP elements in lunar samples provide strong evidence for this \cite{warren_kreep_1979, korotev_kreep_2001, borg_kreep_2004}. These bodies now maintain solid interiors and exhibit various conditions above their surfaces in terms of atmospheric pressure, composition, temperature, and radiation. An exception to this that recent measurements indicate the presence of a liquid silicate layer deep within the interior of Mars \cite{khan_mars_2023}. The evolutionary divergence between Venus and Earth has been suggested to result from the feedback between irradiation, melting, greenhouse forcing and atmospheric escape \cite{2013Natur.497..607H,2021Natur.598..276T}. The vast majority of observed exoplanets orbit within the classical runaway greenhouse instellation threshold, where planets may enter a permanent magma ocean state until their atmospheric volatile inventory is depleted by escape processes \cite{2016SSRv..205..153M,2018SSRv..214...76I}. It has been argued that a more realistic post-runaway steady state may not be compatible with a permanent magma ocean. \citeA{selsis_cool_2023} used radiative-convective models to simulate pure-steam atmospheres under various instellations, finding that the formation of deep isothermal layers allows global radiative equilibrium to be achieved with relatively cool surface temperatures. This indicates that a larger instellation than previously thought may be required to sustain a permanent magma ocean. However, they did not simulate the time-evolution of magma oceans from an initially molten state. Nor did they account for a broad range of opacity sources, such as from gases other than water vapour.
\par 
Understanding the nature of magma ocean evolution is thus of primary importance for resolving the link between a planet's initial conditions and its current state, as it is contemporaneous with stellar evolution and atmospheric escape, as well as providing the initial conditions for potential future geochemical cycling of volatiles through a solidified mantle \cite{2023ASPC..534..907L}. Magma oceans of various depths may occur on planets between giant impact events, influencing the distribution of volatiles and their susceptibility to impact erosion \cite{gu_composition_2024}. Whether generated by stochastic giant impacts or originating from more continuous accretion processes during planet formation, these molten planets are initially self-luminous, rapidly cooling until they reach global radiative equilibrium \cite{2019A&A...621A.125B,salvador_magma_2023,2023ApJ...951L..39K}. Global radiative equilibrium may be sustained by a range of possible conditions; a planet may reach radiative equilibrium before or after complete crystallisation, which determines whether or not the planet can maintain a permanent magma ocean \cite{2021ChEG...81l5735C,2024arXiv240504057L}. The thermal stratification of runaway greenhouse climates has been suggested as an observational signature to probe via exoplanet surveys \cite{2019A&A...628A..12T,2024PSJ.....5....3S}
\par 
Some volatile species (such as \ce{H2O}) are highly soluble in molten rock, allowing significant volatile inventories to be stored within molten interiors \cite{HBO64,sossi_solubility_2020, sossi_solubility_2023, 2023FrEaS..1159412S}. For planets orbiting young stars this could buffer such atmospheres against escape, thereby allowing them to maintain atmospheres for longer, and/or could trap volatiles inside the interior once the planet has solidified \cite{hamano_lifetime_2015,2021ApJ...922L...4D,bower_retention_2022}. The cooling timescale and crystallistion mode of a magma ocean may depend on the thermal blanketing of a radiatively-absorbing atmosphere, since outgassing of sufficiently opaque gases (e.g. \ce{H2O}, \ce{CO2}, and \ce{H2}) could induce a strong enough greenhouse effect to significantly delay cooling of the planet \cite{spada_radius_2013,pierrehumbert_book_2010, 2017JGRE..122.1458S, nikolaou_duration_2019, lichtenberg_vertically_2021, bower_retention_2022}. The composition of the atmosphere overlying a magma ocean is thought to be strongly tied to the properties of the mantle due to the rapid volatile exchange between the two inventories \cite{spaargaren_influence_2020, 2021ApJ...914L...4L}. Given that planetary interiors are expected to span a range of geochemical conditions \cite{2023ASPC..534.1031K}, it is prudent to explore how magma ocean evolution depends on these properties \cite{2021AGUA....200294K,guimond_mineralogical_2023}. 
\par 
In this work, we focus on the impact of interior properties and initial volatile endowment on the evolution of planetary magma oceans in the absence of external loss and delivery processes (e.g. escape). This is done with the aim of understanding how outgassing and energy transport alone shape planetary evolution and solidification outcomes.

\section{Methods}
\label{sec:methods}

\citeA{lichtenberg_vertically_2021} introduced a coupled numerical model of the interior and atmosphere to study the evolution of magma ocean planets, modelling a handful of single-species volatile cases. They found that magma ocean lifetime varies depending on outgassed atmospheric composition, with \ce{H2}-dominated atmospheres yielding long-lived magma oceans, motivating the primary composition of their numerical cases by previously suggested scenarios of planetary evolution. However, they did not link this to a self-consistent description of redox properties. Here, we update their framework of planetary evolution, coupling planetary internal processes to climatic and stellar evolution of evolving magma ocean planets.
\par 
Our one-dimensional column model (`PROTEUS') decomposes the physical system formed by the star and the planet into several components: the interior, the surface, the atmosphere, and the star. The magma ocean is assumed to begin as entirely molten, cooling as energy is lost to space through radiation.

\subsection{Interior}
\label{sec:methods_interior}
\subsubsection{Energy transport}
\label{sec:methods_interior_energy}
The interior component of the system contains the magma ocean and the core. It is solved using an established 1D model (`SPIDER' - \citeA{bower_numerical_2018, bower_linking_2019, bower_retention_2022}). At each level of pressure $p$ and specific entropy $S$, the melt fraction is calculated according to 
\begin{equation}
    \phi = \begin{cases}
            1								& \text{if } S > S_l \\
            (S - S_s)/(S_l - S_s)		& \text{if } S_l \ge S \ge S_s \\
            0								& \text{if } S_s > S
           \end{cases}
    \label{eq:melt_fraction}
\end{equation}
where $S_l$ and $S_s$ are the specific entropies of the liquidus and solidus at $p$. To evolve the interior, SPIDER solves the energy conservation equation in terms of specific entropy,
\begin{equation}
    \int_V \rho \tilde{T} \frac{\partial S}{\partial t} dV = -\int_A \tilde{F} dA + \int_V \rho H dV
    \label{eq:interior_energy}
\end{equation}
where $\tilde{T}$ is the temperature, $\rho$ is the mass density, $\tilde{F}$ is the net heat flux, and $H$ is the internal heat generation. Positive fluxes are measured in the upward direction. $\tilde{F}$ is the sum of contributions by convection, phase mixing, conduction, and gravitational settling. Heat from the radioactive decay of \ce{^40K}, \ce{^232Th}, \ce{^235U}, and \ce{^238U} is included. The initial abundances of these radioisotopes are calculated by scaling measured concentrations for modern Earth ($t=\SI{4.55}{\giga\year}$) to the system's age of the start of the simulations \cite{ruedas_radioactive_2017,turcotte_geodynamics_2002}. The flux boundary condition used to solve the interior model is placed at the topmost interior node. Time-integration of Equation \ref{eq:interior_energy} provides the evolution of the interior's specific entropy, temperature, and melt fraction. The temperature at the top of the interior $\tilde{T_s}$ is used to set the interior constraint on the surface energy balance (Equation \ref{eq:conduction}). The mantle is initialised on an adiabat with a specific entropy $S_0$, which corresponds to an initial surface temperature $>\SI{3000}{\kelvin}$ such that all cases begin with a `hot start' and then cool over time. The core releases heat over time into the bottom of the mantle; it acts as a reservoir of heat set by its initial temperature, density (\SI{10738}{\kilo\gram\per\meter\cubed}), and heat capacity (\SI{880}{\joule\per\kilo\gram\per\kelvin}). We do not model radiogenic heating in the core, nor thermal boundary layers on either side of the core-mantle boundary. This core cooling model is described further in \citeA{bower_linking_2019} and \citeA{lichtenberg_vertically_2021}. Once the base of the mantle reaches the rheological transition, the role of core energetics may become more relevant: during this phase an inner core may begin to crystallise. However, comparatively small changes in the basal temperature of the solid mantle are not the primary focus of this work.
\par 
\subsubsection{Outgassing}
\label{sec:methods_interior_outgas}
Abundances of volatile elements in the melt and gases in the overlying atmosphere are together set by their solubility and equilibrium chemistry. The total masses of \ce{H}, \ce{C}, \ce{N}, and \ce{O} in the planet are prescribed at fixed values, and are partitioned into gaseous volatile species (\ce{H2O}, \ce{CO2}, \ce{N2}, \ce{CH4}, \ce{H2}, and \ce{CO}) as well as dissolved into the melt. The solubility of these volatiles is calculated from empirical fits which depend on the temperature $\tilde{T_s}$ and oxygen fugacity $f\ce{O2}$  \cite{oneill_iw_2002, sossi_solubility_2023, dixon_determination_1995,  libourel_nitrogen_2003, ardia_ch4_2013, armstrong_co_2015}. This accounts for the complex speciation and non-ideal behaviour of volatiles when they dissolve. The three reactions \mbox{\ce{CO2 + 2 H2 <=> CH4 + O2}}, \mbox{\ce{2 CO2 <=> 2 CO + O2}}, and \mbox{\ce{2 H2O <=> 2 H2 + O2}} are assumed to attain thermochemical equilibrium in the atmosphere \cite{schaefer_redox_2017, chase_janaf_1998}. The requirement of elemental mass conservation is used to solve for the volatile partial pressures at the surface using Newton's method. This construction depends on the C/H elemental ratio ($C/H$), total hydrogen inventory (${[}H{]}$), and nitrogen concentration $x_N$ as input parameters to set the total amount of the volatile elements (H, C, N, O) in the atmosphere and magma ocean. Adopting estimates of Earth's primitive mantle, the nitrogen concentration $x_N$ is fixed at 2 ppmw relative to the total mantle mass, although this value could vary between planetary systems \cite{wang_elements_2018}. 
\par 
\subsubsection{Magma ocean solidification}
\label{sec:methods_interior_solidification}
We assume a chondritic mantle based on the solidus and liquidus in \citeA{andrault_mantle_2011} and \citeA{2013Natur.497..607H} --- see \citeA{bower_linking_2019}. Typically, a fully molten mantle with such a composition would freeze out from the bottom to the top, however, even with these assumptions fractional crystallization can induce heterogeneous freeze-out \cite{bower_retention_2022}. This does not account for volatile partitioning into the solid phase. As is a typical assumption in the literature \cite{1993Litho..30..223A,elkins_magma_2012,lebrun_thermal_2013}, we assume that volatiles are homogeneously distributed throughout the melt due to the vigorous convective mixing and stirring from accretion and large impacts. It is possible that mantle convection with large Rayleigh numbers could trap volatile-rich melt within large scale circulations, leading to inhomogeneous volatile distribution and inhibition of outgassing \cite{salvador_magma_2023, siggia_mantle_1994}. Formation of crystals within the solidifying mantle leads to a significant increase in viscosity as the melt fraction passes through the ``rheological transition'' \cite{bower_retention_2022}. The characteristic location of this transition -- the rheological front -- progresses upward during magma ocean solidification. If this occurs rapidly, it could lead to small pockets of melt becoming embedded within the solid part of the mantle, thereby trapping volatiles deep within the planet \cite{mckenzie_intrusion_2011, hier_mantle_2017, elkins_magma_2012}. Together, this means that our outgassing model likely overestimates the amount of volatiles in the atmosphere and therefore the atmospheric surface pressures.

\subsection{Model coupling}
\label{sec:methods_surface}
\begin{figure}[ht]
    \centering
    \includegraphics[width = 0.82\textwidth]{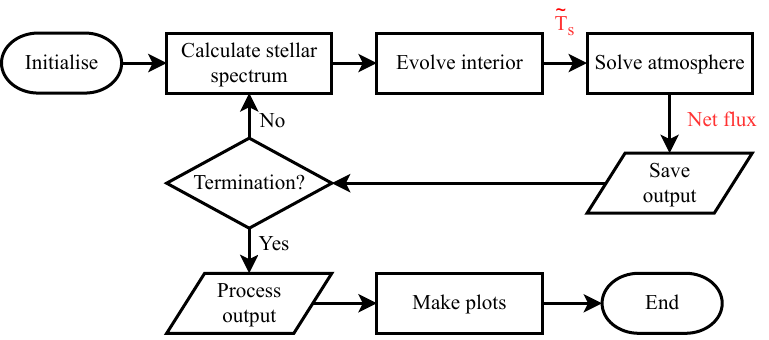}
    \caption{Flowchart outlining our scheme for planetary evolution. The model is initialised into a fully molten state, with the interior and atmosphere coupled together to enable heat transport from the interior into space. At each time-step, the spectral stellar flux impinging on the top of the atmosphere is updated consistently with stellar evolution models, and a check is made for whether Equation \ref{eq:termination_solid} or Equations \ref{eq:termination_steady}a,b are satisfied. The top-of-mantle temperature $\tilde{T_s}$ is calculated by the interior model, subject to the top-of-mantle flux boundary condition derived by the surface conduction (Equation \ref{eq:conduction}) and the outgoing radiation from the top of the atmosphere $F_t$ (Equation \ref{eq:Ft}).}
    \label{fig:proteus_flowchart}
\end{figure}

It is thought that magma oceans could be covered in a thin conductive boundary layer, which limits heat transport through the surface \cite{2007evea.book...91S,schaefer_magma_2016,lebrun_thermal_2013,lichtenberg_vertically_2021}. This boundary layer is analogous to those which form on lava ponds on Earth, but instead occurring on a global or hemispherical scale. This boundary layer is parameterised in our model by a thin layer of constant thermal conductivity $\kappa_c$ and thickness $d_c$ that conducts a heat flux according to Fourier's law
\begin{equation}
    F_c = \kappa_c \frac{\tilde{T_s}-T_s }{d_c}.
    \label{eq:conduction}
\end{equation}
This layer acts to couple the top-of-mantle temperature $\tilde{T_s}$ to the bottom-of-atmosphere temperature $T_s$. The temperature at the top of the boundary layer ($T_s$) is less than that at the bottom $\tilde{T_s}$ (which is fixed by the topmost node of the interior model) for a net positive conductive heat flux $F_c$. Equation \ref{eq:conduction} is used to determine the temperature at the base of the atmosphere by ensuring that $F_c$ is equal to the net (up minus down) bolometric radiative flux at the top of the atmosphere, which is calculated as
\begin{equation}
    F_t = F_{\text{up}}^{\text{LW}} + F_{\text{up}}^{\text{SW}} - F_{\text{down}}^{\text{LW}} - F_{\text{down}}^{\text{SW}}
    \label{eq:Ft}
\end{equation}
This model of surface flux balance is adopted because it conserves energy by construction and allows for temperature discontinuities at the surface. The top of atmosphere radiative flux is used instead of the near-surface flux because radiation will be the primary mechanism for energy transport at the top of the atmosphere, as opposed to at deeper levels where convection may dominate. This is consistent with our stratospheric prescription. \citeA{schaefer_magma_2016} used a similar model for the parameterisation of the boundary layer. The temperature at the bottom of the atmosphere $T_s$ is set by the solution of Equation \ref{eq:conduction}. For a given top-of-mantle temperature $\tilde{T_s}$ (calculated by SPIDER), $T_s$ is determined by using the secant method to find the root of $F_c-F_t$. $F_c$ is the net upward-directed top-of-atmosphere radiative flux calculated using the radiative transfer code described below. This implementation conserves energy between the atmosphere and interior models, but is not guaranteed to do so locally throughout the atmosphere due to the prescriptive nature of the atmospheric temperature profile (Section \ref{sec:methods_atmosphere_profile}). Once the atmospheric fluxes are obtained (Section \ref{sec:methods_atmosphere_radtrans}), $F_t$ is set as the new flux boundary condition at the top of the interior model. This construction allows feedback between the interior and the atmosphere and ensures that energy fluxes are conserved between the components of the model. The flowchart in Figure \ref{fig:proteus_flowchart} shows this graphically.
\par 

\par
We take $\kappa_c = \SI{2}{\watt\per\kelvin\per\meter}$ as a characteristic value of thermal conductivity of this conductive boundary layer, although in reality this quantity would depend on the particular physical and mineralogical properties of the material \cite{balkan_rock_2017, chibati_conductivity_2022}. Previous studies have shown that the thickness of this layer is important for determining the cooling timescale \cite{monteux_cooling_2016}. Observations and models of lava ponds indicate that $d_c$ varies from $\sim \text{cm}$ to $\sim \text{m}$ depending on temperature contrast and flow speed \cite{aufaristama_lava_2018,neri_lava_1998}. Lids on lava ponds are not fully analogous to global magma oceans as the latter cases lack the structural support afforded by the walls of the pools, potentially yielding comparatively smaller $d_c$ \cite{elkins_magma_2012}. Modelling by \citeA{lebrun_thermal_2013} found that $d_c$ varies between $\sim \text{mm}$ and $\sim \text{m}$ during magma ocean crystallisation, until the rheological front reaches the surface, at which point the thickness of the conductive boundary layer increases to $\sim \text{km}$. It is possible that in the later stages of magma ocean evolution that this thick conductive layer could significantly inhibit further cooling. However, it could be destabilised by dynamical stresses on its bottom boundary and impacts on its top boundary \cite{elkins_magma_2012}. The thickness $d_c$ of this layer is set to a constant characteristic value of \SI{1}{\centi\meter} in this work.

\subsection{Atmosphere}
\label{sec:methods_atmosphere}
\subsubsection{Temperature profile}
\label{sec:methods_atmosphere_profile}
The atmosphere plays a key role in determining the rate of cooling of the planet's interior, as thermal emission from the surface will be partially or entirely balanced by downward radiation from the atmosphere and the host star. The composition of our model atmosphere is set by the outgassing calculation described in Section \ref{sec:methods_interior}. We assume that the atmosphere is convective from the planet's surface, in line with previous magma ocean studies \cite{kopparapu_habitable_2013, lichtenberg_vertically_2021, hamano_lifetime_2015, 2021AsBio..21.1325B}. Our atmosphere model (`JANUS') implements the multicomponent non-dilute pseudoadiabat of \citeA{graham_multispecies_2021} by integrating $dT/dp$ from the surface ($p=p_s$) upwards to \mbox{$p = \SI{e-5}{\bar}$}. In regions where all species are in their gas phase, $dT/dp$ is set by the dry adiabat (Equation \ref{eq:dry_adiabat}). 
\begin{equation}
    \frac{dT}{dp} = \frac{1}{c_p(T)} \frac{RT}{p\mu}
    \label{eq:dry_adiabat}
\end{equation}
Moist convection occurs at levels in the atmosphere where any volatile $s$ becomes saturated, which corresponds with a decrease in its mixing ratio. Saturation is defined according to the Clausius-Clapeyron equation. It is assumed that all condensates are instantaneously rained-out and then re-evaporated within dry regions at higher pressures. 
\par 
The gravitational acceleration $g$ and heights $z$ of each level are obtained by simultaneously solving for hydrostatic equilibrium (using the ideal gas equation of state) and Newton's law of universal gravitation; an Euler method is used to integrate these equations from the surface upward. Mean molecular weight $\mu$ [\SI{}{\kilo\gram\per\mole}] and specific heat capacity at constant pressure $c_p(T)$ [\SI{}{\joule\per\kelvin\per\mole}] are obtained from the well-mixed composition and NIST data. 

\subsubsection{Radiative transfer}
\label{sec:methods_atmosphere_radtrans}
The SOCRATES radiative transfer code --- developed by the UK Met Office --- is used to solve the plane parallel two-stream radiative transfer equation \cite{edwards_studies_1996, walters_socrates_2019, sergeev_socrates_2023}. We include gaseous absorption, collisional absorption, and Rayleigh scattering of longwave and shortwave radiation. The exponential sum fitting correlated-k approximation is used for determining the opacity of the gas mixture at each level, depending on its pressure and temperature \cite{amundsen_radiation_2014,amundsen_treatment_2017}. Combined gaseous absorption is treated using the random overlap method \cite{goody_atmospheric_1989,lacis_corrk_1991}. The flexibility afforded by random overlap (with resorting and rebinning of k-terms) can yield smaller errors compared to a reasonably sized pre-mixed k-table \cite{amundsen_treatment_2017}. The k-coefficients are fitted with the number of terms in each band selected such that the root-mean-squared transmission error across each band is less than 1\%. The transmission error is calculated as the difference between the exact transmissions (from the opacity tables) and that calculated using the fitted k-terms and weights.
\par 
We initially calculate k-coefficients directly from the HITRAN2020 database using 318 spectral bands between 0 and \SI{30000}{\per\centi\meter} for comparison with previous work \cite{gordon_hitran_2022}. We also calculate k-coefficients using absorption cross-sections from the University of Geneva's Data and Analysis Center for Exoplanets (DACE) opacity database, tabulated at 80 pressure points (-6 to +3 log bar), 18 temperature points (100 to 2895 K), and 256 bands (1 to \SI{35000}{\per\centi\meter}). This makes our opacity calculations much more complete than previous work; \citeA{kopparapu_habitable_2013} for example tabulated absorption to a maximum temperature of \SI{600}{\kelvin} using 55 bands between 0 and \SI{15000}{\per\centi\meter}, and \citeA{selsis_cool_2023} used 66 spectral bands. Lower transmittances and net fluxes are therefore expected when using the DACE-derived data as it includes more lines (particularly in the optical/UV regimes and at high temperatures). Table \ref{tab:opacity} outlines the included absorbers, their literature sources, and corresponding collisional pairings.
\par 

\begin{table}[ht]
    \centering
    \begin{tabular}{p{0.05\linewidth}  p{0.6\linewidth} p{0.25\linewidth}}
    Gas & References & Collisional pairings \\ 
    \hline
    \ce{H2O} & POKAZATEL -- \citeA{polyansky_H2O_2018}  & \ce{H2O}, \ce{N2}  \\
    
    \ce{H2}  & RACPPK -- \citeA{roueff_H2_2019}  & \ce{H2}, \ce{CH4}, \ce{CO2}, \ce{N2}  \\
    
    \ce{N2}  & WCCRMT -- \citeA{western_N2_2018,western_N2_2017, barklem_N2_2016,shemansky_N2_1969} & \ce{N2}, \ce{H2}, \ce{H2O}  \\
    
    \ce{CO2} & UCL-4000 -- \citeA{yurchenko_CO2_2020}  & \ce{CO2}, \ce{H2}, \ce{CH4}  \\
    
    \ce{CH4} & YT34to10 -- \citeA{yurchenko_CH4_2014, yurchenko_CH4_2017}  &  \ce{CO2}, \ce{H2} \\
    
    \ce{CO}  & HITEMP2019 -- \citeA{li_CO_2015}  & None  \\
    
    \end{tabular}%
    \caption{Sources of line-absorption, literature references, and corresponding collisional absorption pairings. Absorption from the water self-continuum is calculated using Version 3.2 of the MT\_CKD model {\protect\cite{mlawer_mtckd_2012,mlawer_mtckd_2023}}. All other collisional absorption is calculated using the HITRAN collisional absorption database {\protect\cite{karman_hitran_2019}}. 
    The DACE opacity database derives cross-sections using HELIOS-K {\protect\cite{grimm_database_2021}}. 
    }
    \label{tab:opacity}
\end{table}

The downward shortwave flux (absorbed stellar flux - ASF) at the top of the atmosphere is calculated as
\begin{equation}
    F_{\text{ASF}} = F_* f_s (1-\alpha_b) \cos(\theta) 
    \label{eq:asf}
\end{equation}
where $\theta$ is the solar zenith angle, $f_s$ is a scale factor associated with the axial rotation of the planet, $\alpha_b$ captures additional contributions to the planetary albedo, and $F_*$ is the instellation flux (Section \ref{sec:methods_star}). In all cases modelled in this work there is an asynchronous relationship between the planet's orbital period and their day length, so we set $\theta=48.19^\circ$ and $f_s=3/8$ \cite{cronin_choice_2014}. We do not model the impact of clouds on the radiative transfer, so we set $\alpha_b=0$ in all calculations to represent a sky clear of aerosols. Note that $\alpha_b$ does not represent the bond albedo, as Rayleigh scattering and surface reflectivity are modelled as separate and distinct processes. The instellation is calculated as 
\begin{equation}
    F_* = \frac{L_*(t_*)}{4 \pi a^2}
    \label{eq:inst}
\end{equation}
where $L_*(t_*)$ is the stellar luminosity at age $t_*$, and $a$ is the orbital separation.
\par 
Upwelling longwave radiation from the surface of the planet is set by blackbody emission at a temperature $T=T_s$. The surface albedo $\alpha_s$ cannot be tightly constrained due to a lack of experimental data on molten surfaces, however \citeA{essack_albedo_2020} conclude that molten silicate surfaces have a low reflectivity, so we set $\alpha_s$ to $0.1$ in all of our simulations.
\par 

\subsubsection{Stratosphere}
\label{sec:methods_atmosphere_stratosphere}
The assumption that these atmospheres are entirely convective has been challenged. \citeA{selsis_cool_2023} demonstrated that pure-steam atmospheres in global radiative equilibrium may form radiative layers near the surface, which decreases the surface temperature (for a given net flux) compared to a fully-convective atmosphere. It is as-yet unclear how much this extends to atmospheres of more complex composition, and to atmospheres which start in an initially molten and cooling-off state as in this work. Solving for a radiative-convective temperature profile is beyond the scope of this work, but we can make an approximation of a radiative stratosphere -- which is assumed to be transparent to shortwave radiation but a grey absorber of longwave radiation -- by imposing $T(p) \ge T_r$, where 
\begin{equation}
    T_r = \Big( \frac{(1 - \alpha_b) f_s  F_* }{ 2 \sigma_\text{SB}} \Big)^{1/4}
    \label{eq:stratosphere_skin}
\end{equation}
is the radiative skin temperature of the planet, set by the bolometric stellar flux  $F_*$ and the bond albedo \cite{pierrehumbert_book_2010}. $\sigma_\text{SB}$ is the Stefan-Boltzmann constant. This prescription of an isothermal stratosphere does not address the possibility of near-surface radiative layers. We assess whether or not convection shuts off in Section \ref{sec:results_convection} by analysing radiative heating rates throughout the atmospheres of two case studies.

\subsection{Star}
\label{sec:methods_star}
Flux from the host star factors into the evolution of the planet through its radiative energy balance. A star's luminosity evolves both bolometrically and spectrally over time depending on its spectral class and angular momentum endowment -- see Figures 5 and 11 of \citeA{johnstone_active_2021} -- which can impact both the climate and chemistry of orbiting planets \cite{tsai_chemistry_2021, pierrehumbert_book_2010}. We use the model (`MORS') developed by \citeA{johnstone_active_2021} to evolve the stellar spectrum over time self-consistently with the evolution of the orbiting planet. By treating stars as solid shells corotating around solid spheres, MORS models their spin-down sequence and thus obtains X-ray, UV, and bolometric luminosities $L_*$ based on the angular momentum transport between the two rotating bodies, its age $t_*$, and known scaling laws from observations and MHD simulations \cite{johnstone_stars_2017,spada_radius_2013}. Using a star's modern spectrum as a template \cite{gueymard_sun_2003}, we then scale the fluxes within these bands in order to obtain an approximation of its historical emission spectrum. The normalised stellar emission spectrum is updated throughout our simulations at intervals of \SI{0.1}{\mega\year} while the bolometric flux $F_*$ is updated every \SI{200}{\year}. For this work, we treat the Sun as an `intermediate' rotator by placing it in the 60th percentile for rotation speed for its mass and age \cite{gallet_stellar_2013}.

\subsection{Model termination}
\label{sec:methods_termination}
The model terminates under two possible phase states at the end of the simulation. This occurs either when the planet has solidified (where the global melt fraction $\Phi$ is less than a critical value)
\begin{equation}
    \Phi < 0.005,
    \label{eq:termination_solid}
\end{equation}
or when the planet enters into a steady state (where the global melt fraction is unchanging and the net radiative flux $F_t$ is small -- cf. Equation \ref{eq:Ft}).
\begin{subequations}
    \begin{align}
        |d\Phi/dt| &< \SI{e-10}{\per\year} \\
        F_t        &< \SI{0.8}{\WPMS}
    \end{align}
    \label{eq:termination_steady}
\end{subequations}
This means that the lowest possible surface temperature in our model is set by the solidus temperature of the mantle. As a result, the surface temperatures are always supercritical and no oceans are formed upon rainout. It is possible that these planets would continue to cool post-solidification, and potentially form liquid oceans above a solidified mantle. Figure \ref{fig:proteus_flowchart} outlines the coupling process by which the components of the system are able to interact. The following subsections outline each component in detail.

\subsection{Experimental configuration}
We run our model across a grid of parameters to explore their impact on the composition of the atmosphere, the phase state at model termination (i.e. a solid or molten mantle), and the duration of the rapid cooling stage. The grid is configured according to the parameters in Table \ref{tab:grid}. All cases set $M_p = M_\oplus$, $R_p = R_\oplus$, and $R_{\text{core}} = 0.55 R_p$.

\begin{table}[ht]
    \centering
    \begin{tabular}{llll}
    Parameter  & Symbol & Points & Range or values \\ 
    \hline
    Orbital separation       & $a$      & 7 & 0.105, 0.316, 0.527, 0.737, 1.054, 2.108, 3.162  \\
    
    Oxygen fugacity          &  $f$O$_2$  & 7 & -5, -3, -1, 0, 1, 3, 5  \\
    
    C/H ratio                & C/H      & 7 & 0.01 to 2.0 logarithmically spaced \\
    
    Total H inventory & {[}H{]}  & 3 & 1, 5, 10 \\
    
    \end{tabular}%
    \caption{Parameter axes used to construct our grid of models. There are 1029 grid points in total. Orbital separation is measured in AU from the Sun (Equations \ref{eq:asf} and \ref{eq:inst} in Section \ref{sec:methods_atmosphere_radtrans}). The other three parameters relate to the outgassing (Section \ref{sec:methods_interior_outgas}) of volatiles: oxygen fugacity $f\ce{O2}$ is measured in \mbox{log$_{10}$} units relative to the Iron-W\"ustite (IW) buffer, C/H is the ratio of the total mass of carbon to the total mass of hydrogen in the mantle and atmosphere combined, the total hydrogen inventory ${[}H{]}$ is measured in units of the total amount of hydrogen in all of Earth's oceans (hereafter abbreviated to `oceans').}
    \label{tab:grid}
\end{table}
\par  
The range of $a$ bounds the habitable zone (roughly 0.95 to 2.4 AU) and extends to small separations for applicability to Venus and highly irradiated exoplanets \cite{kasting_hz_1993,ramirez_hz_2017}. Bulk elemental endowment depends on the location and timing of planetary formation \cite{2023ASPC..534..717D,2023ASPC..534.1031K}, so we use a broad range of $[H]$ and C/H which encompass concordant estimates for primitive Earth \cite{wang_elements_2018}. Mineralogical variations alone will introduce a range of at least 4 log units to $f\ce{O2}$ \cite{guimond_mineralogical_2023}, however, inferences of the redox state of material accreted onto white dwarf stars indicates an oxygen fugacity comparable to asteroids within the solar system \cite{doyle_white_2019}. Constraints on Mars' redox state derived from measurements of martian meteorites imply a variation of at least 3 log units \cite{2001Sci...291.1527W}. Observations of Mercury's surface by the MESSENGER spacecraft have been used to infer a highly reduced composition \cite{cartier_reducing_2019}. We therefore explore a range of $f\text{O}_2$ from extremely reduced to oxidised.

\section{Results}
\label{sec:results}

\subsection{Model validation}
\label{sec:results_validation}
\begin{figure}[ht]
    \centering
    \includegraphics[height=2.25in]{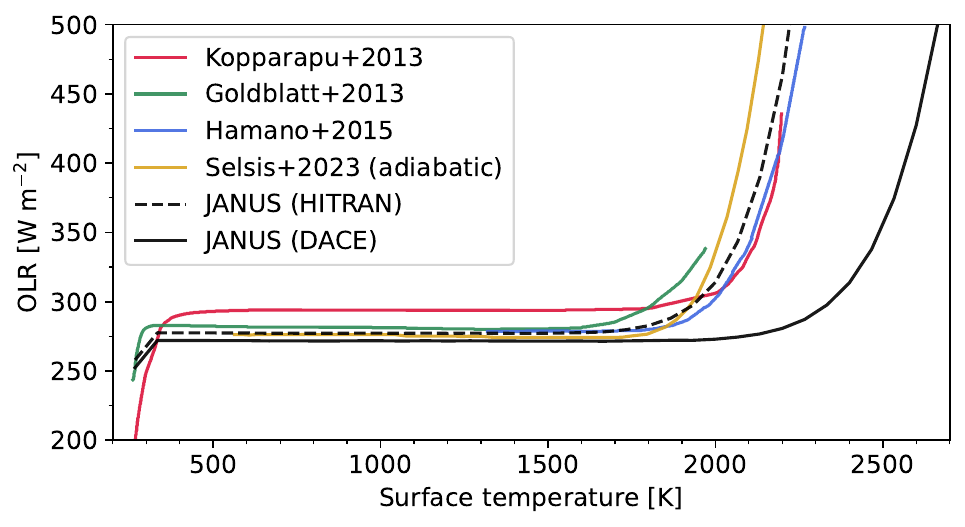}
    \caption{OLR versus surface temperature for thick pure-steam atmospheres. JANUS is compared against curves from {\protect\citeA{hamano_lifetime_2015}}, {\protect\citeA{goldblatt_low_2013}}, {\protect\citeA{kopparapu_habitable_2013}}, and {\protect\citeA{selsis_cool_2023}}. Taking the value at \SI{1400}{\kelvin}, we find a runaway OLR limit of \SI{277}{\WPMS} using HITRAN-derived opacities (dashed black line), which is close to the canonical Simpson-Nakajima limit of \SI{280}{\WPMS} {\protect\cite{komabayasi_limit_1967}}. The solid black line plots values calculated with JANUS using DACE-derived opacities, which yields an OLR limit of \SI{272}{\WPMS}. With the most complete absorption data, the post-runaway transition shifts to higher temperatures ($\sim 2100 \text{ K}$ as opposed to $\sim 1800 \text{ K}$).}
    \label{fig:runaway}
\end{figure}

A key requirement is that our atmosphere model is able to reproduce the runaway greenhouse effect of a pure-steam convective atmosphere. The temperature structures of our model atmospheres are directly tied to the surface temperatures and pressures; each layer can be  in one of three states: on the dry adiabat, on a moist pseudoadiabat, or isothermal if above the tropopause. For sufficiently large $T_s$, most of the atmosphere will lie on the dry adiabat (Equation \ref{eq:dry_adiabat}). At cooler $T_s$, some fraction of the column will be condensing in which $T(p)$ is set only by the condensing gas through the Clausius-Clapeyron relation. Therefore, the outgoing longwave radiation (OLR) becomes decoupled from $T_s$ in cases where the effective photosphere is within a condensing region, leading to the so-called `runaway OLR limit'. Only when the deep atmosphere (and/or surface) are sufficiently hot can the planet enter into a post-runaway regime, where the photosphere is within a dry convective region which couples $T_s$ to the OLR \cite{goldblatt_low_2013,boukrouche_runaway_2021,innes_runaway_2023}. Additionally, since emission from the atmosphere and surface is set by the Planck function, hotter temperatures lead to emission at shorter wavelengths, allowing for increased transmission through windows between \ce{H2O} absorption features and OLR to increase with $T_s$\cite{pierrehumbert_book_2010}. This behaviour is important because it limits how fast the planet may cool. In order to test the radiative transfer, Figure \ref{fig:runaway} plots OLR versus surface temperature. JANUS is demonstrated to compare well against previous studies when calculating fluxes using the HITRAN-derived opacities. Due to increased line absorption below \SI{500}{\nano\meter}, the calculations using DACE-derived (POKAZATEL Version 2) opacities finds a lower OLR limit and a hotter transition into the post-runaway regime. This is because higher $T_s$ are required to emit through the comparatively narrower windows, and increased opacity shifts the photosphere to lower pressures. Lower temperatures yield minimal differences in the flux between the two linelists, because HITRAN is sufficiently complete in this regime. 

\par 
Given the large difference between the two JANUS cases in Figure \ref{fig:runaway}, we further validate the DACE-derived opacities and our pipeline for processing them. Firstly, we calculate k-coefficients for absorption by pure \ce{O2} gas using opacities from DACE, which are themselves derived from the HITRAN linelist for \ce{O2}. There are minimal differences ($\sim 0.16\%$) in OLR calculations with these k-coefficients compared to when using k-coefficients and opacities calculated from the HITRAN linelist directly (not from DACE). This indicates that our k-coefficients derived from DACE are accurate. When fitting k-coefficients to a tighter tolerance of 0.0005\%, there are minimal differences in the OLR (< \SI{1}{\WPMS} at \SI{2800}{\kelvin}). We also perform a sensitivity test on the number of spectral bands by calculating the OLR of a pure-steam atmosphere with $T_s = \SI{2800}{\kelvin}$: compared to calculations with 4096 bands, there is a 1.8\% decrease in OLR with 256 bands, a 14.2\% decrease with 48 bands, and a 61.7\% decrease with 16 bands. This trend can be explained by narrow emission windows being better resolved with more bands \cite{2017JGRE..122.1458S}. 256 bands are used for the rest of this work. The continua databases used in this work were developed for use alongside HITRAN, so we also test whether the aforementioned flux differences (between simulations using DACE- or HITRAN-derived opacities) could be attributed to continuum incompatibility with the linelists we used. In line with the predictions of \citeA{mlawer_mtckd_2023}, we find that the relative flux differences are present even with the continua disabled, meaning that they can be attributed to differences in line absorption.
\par
\begin{figure}[ht]
    \centering

    \includegraphics[width=\textwidth]{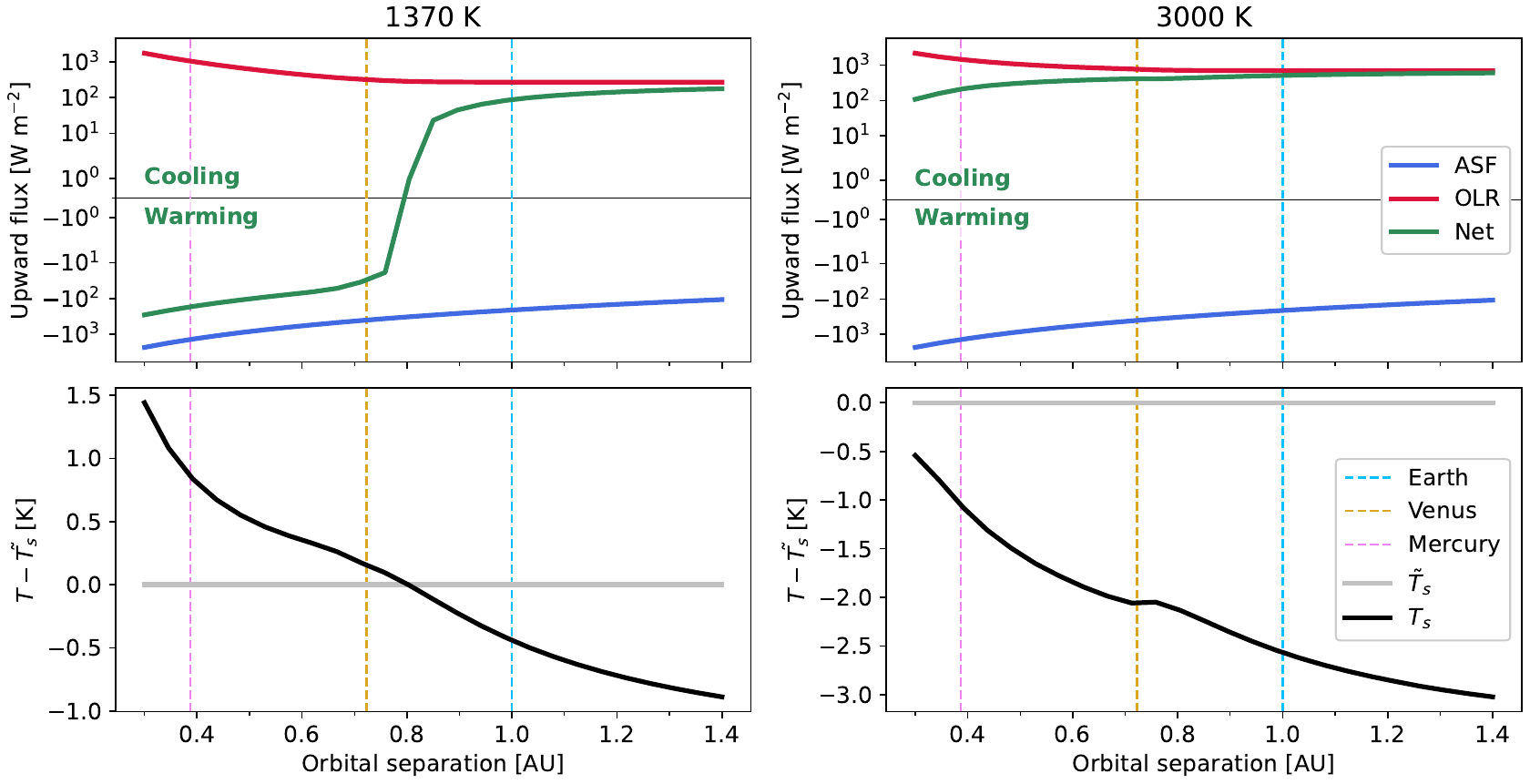}
    
    \caption{Global energy balance versus orbital separation for two cases of mantle temperature \SI{1370}{\kelvin} (\textbf{left}) and \SI{3000}{\kelvin} (\textbf{right}). The top panels plot radiative fluxes versus orbital separation. Absorbed stellar flux (ASF) is the downward top of atmosphere shortwave radiation flux (Equation \ref{eq:asf}). Green lines plot the net radiative flux at the top of the atmosphere $F_t$. The bottom panels show the surface temperatures $T_s$ (black line) required to maintain the fluxes in the upper panels, relative to the mantle temperatures $\tilde{T_s}$ (grey line).}
    \label{fig:conduction}
\end{figure}

 
Bodies within the solar system provide a ground truth against which to calibrate and test planetary models. One key requirement is that a model of Earth should solidify, while highly irradiated planets should maintain a permanent magma ocean. To this end, we solved for the radiative fluxes across a series of different orbital separations from a young Sun ($t_* = \SI{100}{\mega\year}$), setting the planet radius and mass to that of Earth, and prescribing a pure-steam atmosphere of \SI{280}{\bar}. Figure \ref{fig:conduction} plots the resultant fluxes for $\tilde{T_s} = \SI{1370}{\kelvin}$ (left, representing a solid mantle from \citeA{hamano_lifetime_2015}) and \SI{3000}{\kelvin} (right, representing a completely molten mantle). This demonstrates that a model of Earth would cool below the surface solidus even with a thick steam atmosphere in both cases. At Venus' orbital separation with $\tilde{T_s} = \SI{3000}{\kelvin}$, the planet emits a positive net top of atmosphere flux which indicates that it would also initially cool. Its net negative top of atmosphere flux for $\tilde{T_s} = \SI{1370}{\kelvin}$ indicates that it would reach radiative equilibrium before cooling to that surface temperature, thereby maintaining some amount of melt under a \SI{100}{\bar} steam atmosphere. A planet at Mercury's orbital separation behaves similarly to that at Venus' in both cases, maintaining some level of melt. These behaviours are validated by the knowledge that Earth no longer maintains a magma ocean, and that Venus may have maintained one until a sufficient amount of water had escaped and/or the irradiation from the Sun decreased \cite{salvador_magma_2023, 2009E&PSL.286..503G, 2013Natur.497..607H}. These results demonstrate that -- within the given parameters -- this formulation for the surface coupling of the interior and atmosphere components of the system behaves reasonably. 
\par 
The radiative transfer component of our radiative-convective model does not require further validation here, as it has been well tested under a range of conditions \cite{boukrouche_runaway_2021, innes_runaway_2023, sergeev_socrates_2023, lichtenberg_vertically_2021}. We also verified that our integration of the hydrostatic, ideal gas, and gravity equations is reasonable. By comparing against data retrieved from Earth-observing satellites, we found that the Euler integrator provides good accuracy using 210 levels \cite{dudhia_rfm_2017}.

\subsection{Fiducial pure-steam atmosphere}
As a fiducial case, a test planet is initialised as fully molten and endowed with an arbitrary 10 bar pure-steam atmosphere. The total water endowment (equivalent to $\text{[H]} = 8.5$ times that in Earth's oceans) is therefore set by the solubility law and this initial pressure. The planet is placed around a young Sun ($t_* = \SI{100}{\mega\year}$) at 1 AU. We set $M_p = M_\oplus$, $R_p = R_\oplus$ which approximately represents the configuration of a young Earth.
\par
\begin{figure}[ht]
    \centering
    \includegraphics[width=\textwidth]{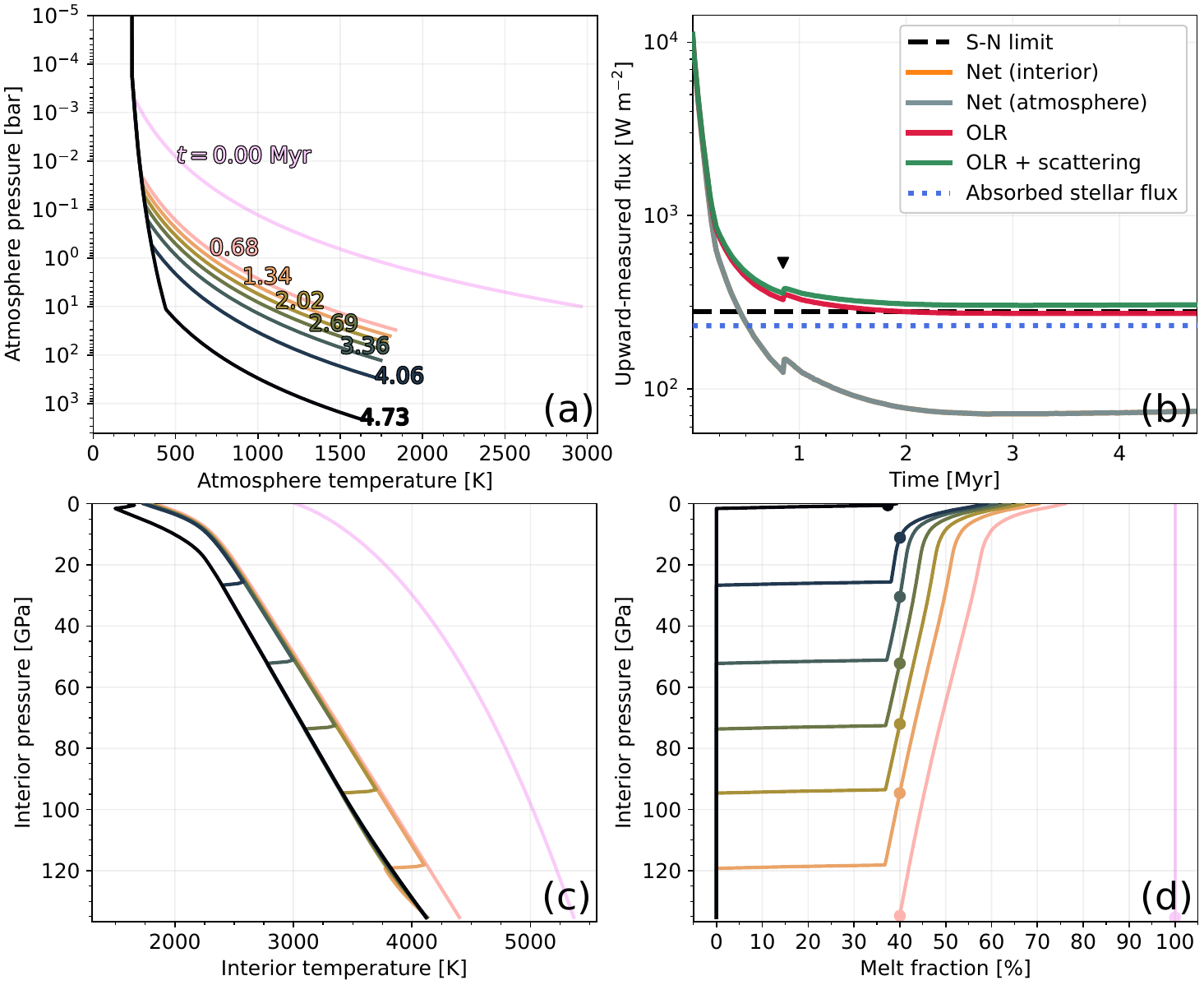}
    \caption{Time evolution of the fiducial model until solidification occurs. Panel a: atmosphere temperature profiles, with simulation time [Myr] indicated by the coloured text on each curve. Panel b: upward-measured energy fluxes, with the Simpson-Nakajima limit indicated by the dashed black line. Panel c: interior temperature profiles. Panel d: interior melt fraction profiles, with the rheological front ($\phi=40\%$) on each profile indicated by a dot. }
    \label{fig:fiducial_results}
\end{figure}

 
The surface temperature is initially very large ($T_s = \SI{3023}{\kelvin}$) which places the planet into a post-runaway state that enables large net outgoing fluxes ($F_t = \SI{10.6}{\kilo\WPMS}$). The planet cools rapidly, resulting in a solidification time of \SI{4.7}{\mega\year}. This is too short for significant stellar evolution to occur \cite{baraffe_new_2015}. Figure \ref{fig:fiducial_results} shows the evolution of this planet's atmosphere and interior temperatures, energy fluxes, and mantle melt fraction. Panel a shows that the surface temperature of \SI{1615}{\kelvin} at solidification is within the runaway greenhouse regime indicated by Figure \ref{fig:runaway} due to the region of deep moist convection between \SI{100}{\bar} and \SI{0.01}{\bar}. The simulation spends most of its time in the runaway greenhouse regime which limits the OLR to \SI{273}{\WPMS}, which, minus the incoming stellar radiation and the scattering contribution, results in a net outgoing flux of \SI{74}{\WPMS} (panel b). The planet does not attain radiative equilibrium in these simulations because solidification occurs first. The non-zero net upward energy flux at solidification (panel b) indicates that the planet would continue to cool below the surface solidus temperature. The interior cools rapidly from its initial state to below the liquidus (panel c). It then spends a comparatively longer period with a mushy mantle until solidification occurs (panel d). Panel d also shows the ascent of the rheological front (defined at $\phi=40\%$ as in \citeA{lebrun_thermal_2013}), which tracks where the mantle undergoes a rheological transition between dynamics characteristic of a melt to those of a solid \cite{bower_retention_2022}. The black triangle in panel b indicates the time at which the rheological front begins its ascent (cf. panel d), at which point mixed-phase energy transport processes enter into the system.
\par 
The grey and orange curves in Figure \ref{fig:fiducial_results}b overlap continuously, demonstrating that our method for loosely coupling the interior and atmosphere components of the model is reasonable. The contribution from Rayleigh scattering is small (\SI{34}{\WPMS}) but not insignificant compared to the OLR at solidification. This result is consistent with Section \ref{sec:results_validation} in that an Earth-like case solidifies under a pure steam atmosphere. The isothermal stratosphere (\SI{235}{\kelvin}) has a small presence although it has little impact on the simulation due to the small opacity at such low pressures. 
\par 
The solidification time of \SI{4.7}{\mega\year} is comparable with the value of \SI{3.9}{\mega\year} found by \citeA{hamano_lifetime_2015} for a planet at 1 AU. The \SI{0.8}{\mega\year} difference may be reasonably accounted for by different total reservoirs of hydrogen and oxygen compared to our case, different opacities and melting curves, and our inclusion of Rayleigh scattering and non-zero surface albedo. \citeA{lebrun_thermal_2013} found that magma ocean duration is ``quite sensitive'' to the planet's \ce{H2O} endowment due to the large radiative opacity of \ce{H2O} gas. Similarly, \citeA{hamano_lifetime_2015} found that the solidification time of an Earth-like lava planet strongly depends on its initial water inventory, increasing from 0.7 to \SI{630}{\mega\year} for 0.1 and 10 oceans respectively (their Figure 4b), the latter case only being able to cool due to atmospheric escape. \citeA{lebrun_thermal_2013} also found that the magma ocean duration could vary in length by between 0.8 and \SI{2}{\mega\year} depending on the choice of mantle solidus and liquidus curves; this factor being particularly important when comparing our results with the literature given that PROTEUS is able to model the magma ocean in 1D. Previous works used parameterised adiabatic mantles, which assume that convection is the only energy transport process (effectively making them zero-dimensional). With these modelling differences under consideration our solidification time is taken to be reasonable. Solidification of the mantle leads to significant outgassing of water, with the surface partial pressure monotonically increasing from 10 to \SI{2054}{\bar} in this simulation.
\par 
Solidification is bottom-up, so the surface layer is the last to begin to solidify (Figure \ref{fig:fiducial_results}d). This occurs at \SI{0.2}{\mega\year}, at which point crystals are forming within the melt and settling at the bottom of the magma ocean \cite{bower_retention_2022, solomatov_suspension_1993, solomatov_nonfractional_1993, solomatov_fluid_2000}. While convection rapidly transports energy upwards ($\sim \SI{e10}{\WPMS}$), this is largely offset by the effective downward transport of energy by latent heat ($\sim -\SI{e10}{\WPMS}$). Additionally, gravitational settling acts to transport energy upwards ($\sim \SI{e7}{\WPMS}$). Inclusion of these additional energy transport processes alongside convection reduces the net upward energy transport, and correspondingly increases the magma ocean cooling time. Although we include radiogenic heating for completeness, the simulated global heat flux through the surface of the planet arising from radioactive decay is initially \SI{0.282}{\WPMS} and then decreases by 0.267\% over the course of the simulation. This is negligible compared to the other heat transport processes in this case.

\subsection{Parameter exploration}
\label{sec:results_grid}

\begin{figure}[ht]
    \centering
    \includegraphics[width=0.999\textwidth]{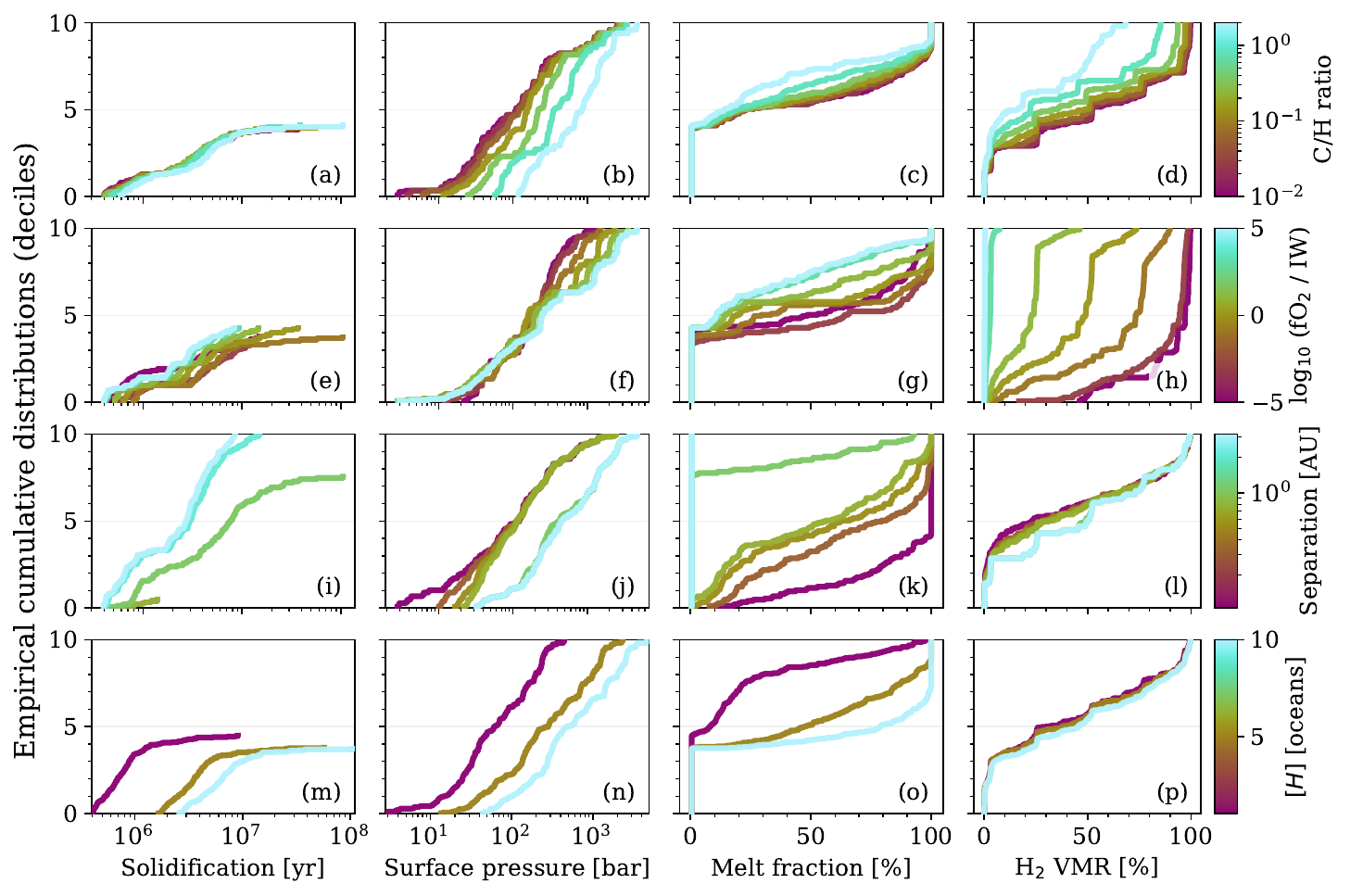}
    \caption{Empirical cumulative distribution functions for four independent variables (colourbars) and four dependent variables (x-axes). Curves are produced for an independent/dependent variable combination ($v_i$, $v_d$) by filtering grid cases to each value of $v_i$ (rows), extracting the corresponding $v_d$ (columns) values at model termination, and then sorting the $v_d$ values in ascending order. For a given curve the corresponding $v_i$ is held constant while all other independent variables are allowed to vary, thereby producing the distribution. If a variable $v_d$ has a strong dependence on a variable $v_i$, then the distributions in the corresponding panel will tend not to overlap. Steep curves (in the $dy/dx>0$ sense) correspond to $v_d$ being distributed around a small range of values, while shallower curves correspond to a wider distribution. The data are plotted for all four axes of the parameter space (Table \ref{tab:grid}) and a selection of four dependent variables (solidification time, total surface pressure, interior melt fraction, and \ce{H2} mole fraction). Cases which reach radiative equilibrium are treated as having an infinite solidification time, and thus do not appear explicitly in panels a, e, i, and m.}
    \label{fig:ecdf_many}
\end{figure}

Across our parameter grid of 1029 cases, 413 (40\%) result in complete solidification, while 616 (60\%) reach radiative equilibrium and thereby retain permanent magma oceans of various depths. The minimum, median, and maximum times to solidify are \mbox{\SI{4.0e5}{}}, \mbox{\SI{3.0e6}{}}, and \mbox{\SI{1.1e8}{}} yr respectively. The maximum time for a case to reach radiative equilibrium is  \SI{1.4e8}{} yr. Figure \ref{fig:ecdf_many} plots distribution functions for all cases explored, which statistically represents how four dependent variables (columns) vary according to the four independent variables (rows, Table \ref{tab:grid}). A broad range of evolution pathways are demonstrated to be possible for a given planet, even at a fixed orbital separation.
\par 
The solidification time is sensitive to the orbital separation (panel i), typically decreasing with orbital distance. This is because the incoming stellar flux is smaller at larger distances, so the outgoing flux becomes increasingly dominant in the total energy balance. The solidification time is shortest (median value approx. 3 Myr) at 3 AU, where the planet rapidly cools and solidifies. As demonstrated by the fiducial case, complete solidification leads to significant outgassing (panel j) and large surface pressures. The longest solidification time (approx. 90 Myr) is larger than that found by the fiducial pure-steam case, indicating that the lifetimes of non-permanent magma oceans may be extended by atmospheres of mixed compositions due to the additional sources of opacity. The global melt fraction is directly tied to the orbital separation (panel k) such that less irradiated cases more commonly solidify. Complete solidification only occurs for a small fraction of cases for $a \le 0.7 \text{ AU}$ (panel i). 
\par
Across all cases and times across our grid of simulations, the surface pressure is maximised at \SI{4711}{\bar} by a \ce{H2O}- and \ce{CO2}-dominated atmosphere produced by a solidified magma ocean after 4.7 Myr of evolution. Some non-solidifying cases attain global radiative equilibrium very slowly due to the imbalance of outgoing and incoming radiation being very small. The most extreme example of this (case 635) evolved with a thick \ce{H2} dominated atmosphere for 142 Myr before satisfying our criteria for radiative equilibrium (Equation \ref{eq:termination_steady}).
\par 
Alongside the orbital separation, the hydrogen inventory of the planet $[H]$ exerts significant control over the solidification time (panel m). The solidification time increases with $[H]$. The oxygen fugacity and C/H ratio have a smaller but non-zero impact on the solidification time (panels a and e). The ensemble-median melt fraction varies from 10\% to 80\% depending on the value of $[H]$ (panel o). $[H]$ is critical to partially molten cases because speciation of hydrogen atoms into \ce{H2}, \ce{H2O}, and \ce{CH4} upon outgassing contributes significant radiative opacity to the atmosphere, slowing energy loss to space and thereby prolonging the evolution of the planet and potentially preventing solidification entirely. 
\par
\begin{figure}[ht]
    \centering
    \includegraphics[width=\textwidth]{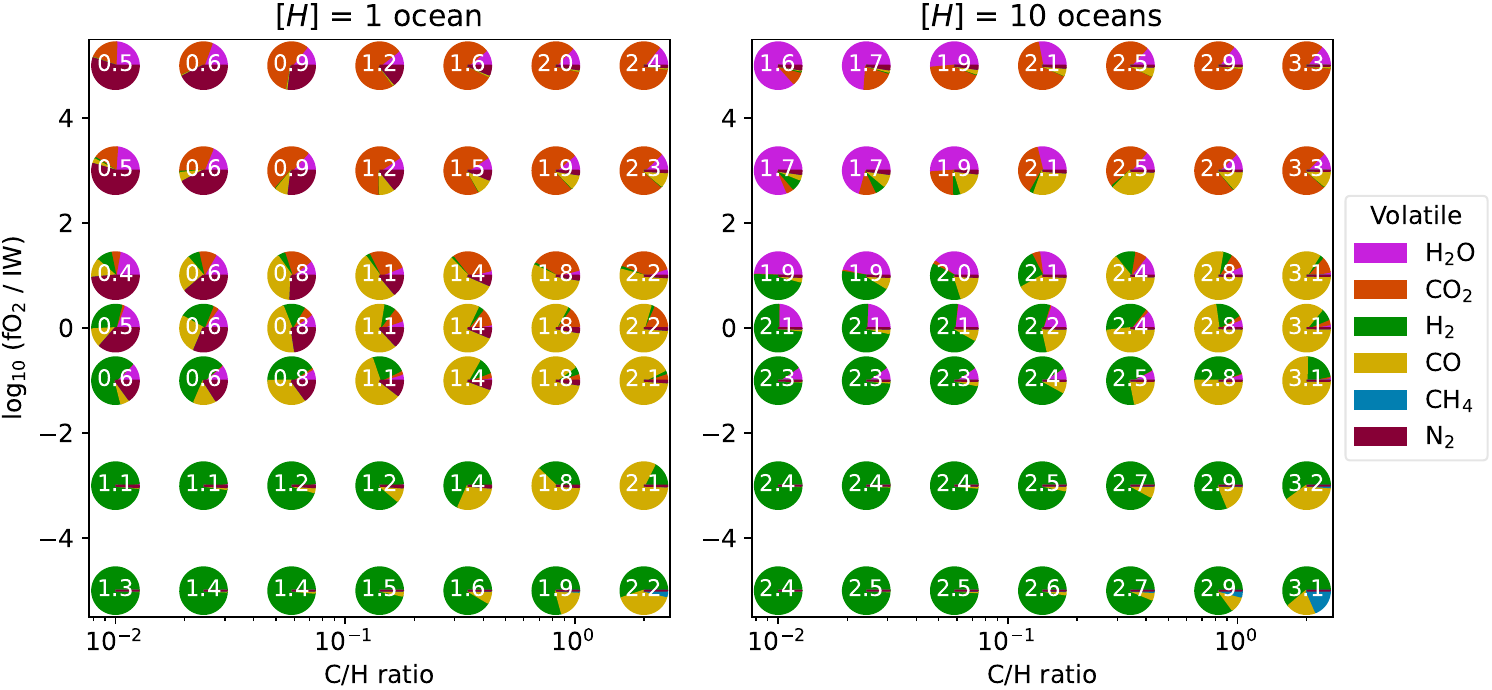}
    \caption{Atmospheric composition for planets which do not solidify. Pie charts display the atmosphere mole fractions $\chi_i$ (equivalently, volume mixing ratios $n_i/n_{\text{total}} = p_i/p_{\text{total}}$) for all six included volatiles across the parameter space at \SI{0.1054}{\AU}. The white numbers enumerate the total surface pressure as $\text{log}_{10}(p_s/\text{bar})$, rounded to 1 decimal place. Note that mole fraction is a relative measure of molecule number density, so even at $\text{C/H}=2$ (a measure of mass ratio) it is possible for there to be more \ce{H2O} than \ce{CO2}.}
    \label{fig:molten_pies}
\end{figure}

 
There is a large range of \ce{H2} mole fraction across the parameter space (panel h), with it primarily being controlled by the oxygen fugacity of the mantle. \ce{H2} is consistently dilute (maximum mole fraction <5\%) under the most oxidising conditions ($\text{fO}_2 \ge \text{IW}+3$), but is consistently dominant (median mole fraction >50\%) in reducing cases ($\text{fO}_2 \le \text{IW}-1$; see also Figure \ref{fig:molten_pies}). This is because larger $f\text{O}_2$ promotes the outgassing of oxygen-bearing molecules (such as \ce{H2O}) over unoxygenated ones (such as \ce{H2}), with variability depending on the gas thermochemistry. Since the model terminates at solidification or radiative equilibrium, surface temperatures always remain high and the abundances of less themochemically-stable reduced species (e.g. \ce{CH4} -- \citeA{chase_janaf_1998}) remain small. The production of \ce{H2} dominated atmospheres under reducing conditions are inline with the findings of \citeA{schaefer_redox_2017} and \citeA{ortenzi_redox_2020}.
\par 
While \mbox{C/H} has only a small impact on the solidification state of the planet (panels a and c), it does yield changes to the total surface pressure (panel b). A large \mbox{C/H} promotes carbon-bearing molecules (such as \ce{CO2}) over hydrogenated species (such as \ce{H2}, panel d), resulting in significant changes to the mean molecular weight $\mu$ of the atmosphere. Surface pressure increases linearly with $\mu$ through the hydrostatic equation, yielding higher pressures at larger \mbox{C/H}. This has less impact in solidified cases where outgassing is complete and the total pressure is large, where the atmosphere is typically dominated by steam. Since $\mu$ also controls the scale height of the atmosphere, constraints on bulk C/H could be made from observations of permanently molten planets. At the largest \mbox{C/H} the partial pressure of \ce{CO2} approaches 1800 bar, but \ce{H2O} typically remains a dominant or major component of the atmosphere.
\par 
Figure \ref{fig:molten_pies} explicitly plots the atmospheric composition for highly-irradiated planets across two slices of the parameter space (1 ocean and 10 oceans of hydrogen). While the atmospheric composition depends on $f\text{O}_2$ and C/H, $[H]$ exerts the most control over the dominant gas. \ce{CO}- and \ce{CO2}-dominated atmospheres are produced for $\text{C/H} > 0.1$ and $\text{fO}_2 \gtrsim \text{IW}-1$, inline with \citeA{bower_retention_2022} and \citeA{sossi_solubility_2020}. \ce{H2}- and \ce{CO}-dominated atmospheres are common, particularly under reducing conditions, inline with the results of \citeA{maurice_volatile_2024}. The mole fraction of \ce{H2O} is typically small across all cases plotted in Figure \ref{fig:molten_pies} because it is primarily dissolved into the molten mantles. Methane remains a minor component in all of these cases due to its thermochemical instability and its solubility, with its mixing ratio maximised for the most reducing and hydrogen-rich cases. The low molecular weight of \ce{H2} means that a large partial pressure is required to satisfy the hydrogen mass requirement in cases with large hydrogen inventories, and hence the wide range of cases with \ce{H2} dominated atmospheres. The total surface pressure $p_s$ varies by approximately 2 dex in both cases, with larger $p_s$ typically found for larger $[H]$ and \mbox{C/H}. There is a broader range of $p_s$ at oxidising conditions than at reducing conditions, corresponding to a larger diversity of composition. 

\par
\begin{figure}[ht]
    \centering
    \includegraphics[width=\textwidth]{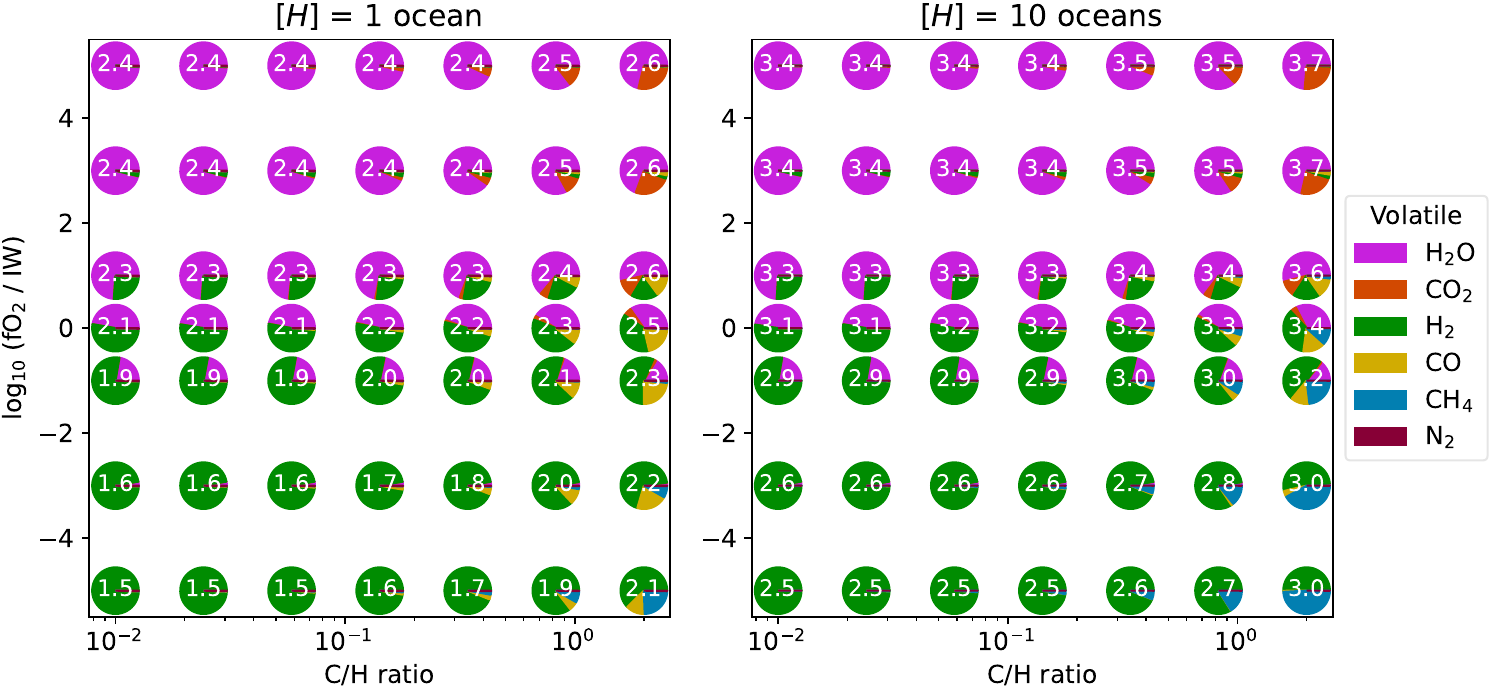}
    \caption{Same as Figure \ref{fig:molten_pies}, but for cases at \SI{3.162}{\AU}, which are solidified.} 
    \label{fig:solid_pies}
\end{figure}

 
Figure \ref{fig:solid_pies} plots atmospheric composition for two slices of the parameter space at \SI{3.162}{\AU}, thereby probing the atmospheres of planets which have solidified. The diversity of composition is smaller than for molten cases (Figure \ref{fig:molten_pies}). This is primarily because the high solubility of hydrogen-bearing volatiles leads to a large fraction of the hydrogen being dissolved in the magma ocean in molten cases, whereas in these cases it is primarily partitioned into the atmosphere because the volume of melt is small. Lower surface temperatures also disfavour the formation of \ce{CO}, decreasing its abundance relative to \ce{CO2} in these cases. These atmospheres typically have higher surface pressures than in the molten cases of Figure \ref{fig:molten_pies} because of the outgassing that occurs with solidification. Because the majority of volatiles have been outgassed, there is a much smaller range of $P_s$ in these cases, which instead set primarily by $[H]$, $f\text{O}_2$, and \mbox{C/H} and not the variable melt fraction.

\subsection{Dihydrogen and water}
\label{sec:results_hydrogen}
\begin{figure}[ht]
    \centering
    \includegraphics[width=\textwidth]{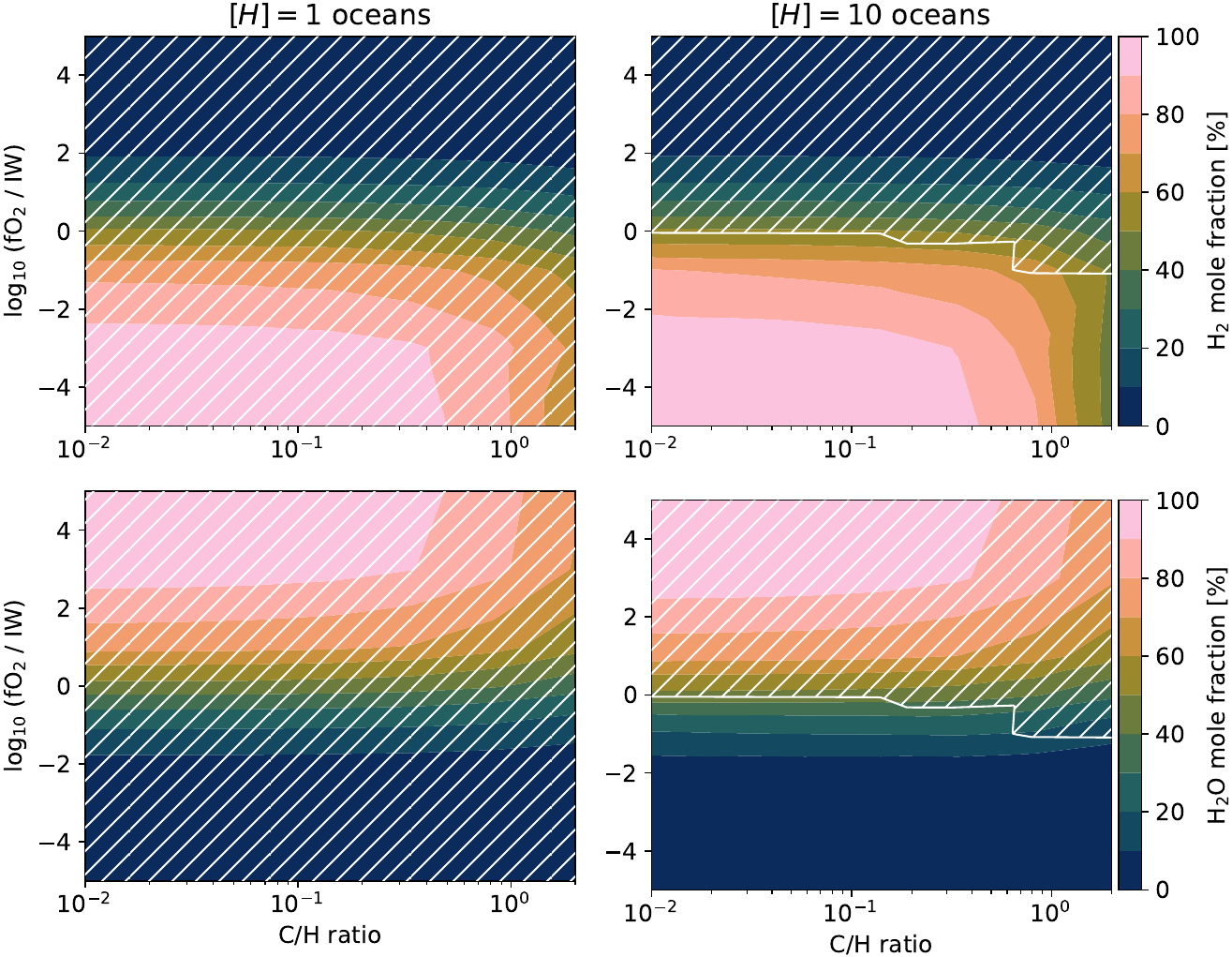}    
    \caption{
    Mole fractions of \ce{H2} and \ce{H2O} (rows) plotted versus oxygen fugacity (y-axes) and C/H (x-axes), for two values of planetary hydrogen inventory $[H]$ (columns). The white hatched regions indicate fully solidified cases. Separations of 1.054 AU.
    }
    \label{fig:hydrogen_water}
\end{figure}

Figure \ref{fig:ecdf_many} shows that the total hydrogen inventory and redox state of a magma ocean planet exerts control over its evolution but does not explain the mechanism behind this relationship. Vibrational modes within triatomic \ce{H2O} and collisional absorption of \ce{H2} make both of these molecules effective greenhouse gases \cite{pierrehumbert_book_2010, pierrehumbert_greenhouse_2011, borystow_cia_1991, abel_cia_2011, chen_greenhouse_2011}. Similarly, \citeA{nikolaou_duration_2019} and \citeA{2019ApJ...875...31K} found that planetary volatile endowment is linked to magma ocean cooling and spectral response through outgassing of greenhouse gases. Figure \ref{fig:hydrogen_water} plots the mole fractions of \ce{H2} and \ce{H2O} for cases at \SI{1.054}{\AU}, thereby probing the parameter subspace where either solidification or radiative equilibrium outcomes are probable. All cases solidify for $[H] = 1 \text{ oceans}$, independent of the oxygen fugacity and C/H ratio (left panels). The top-right panel shows that molten cases have an increased mole fraction of \ce{H2}, affirming the indications of Figure \ref{fig:ecdf_many} that the \ce{H2} greenhouse effect is key to preventing/slowing planetary solidification.
\par 
For $[H] = 10 \text{ oceans}$ the phase state is more sensitive to C/H and $f\text{O}_2$, although comparison of the two columns indicates that $[H]$ still exerts the most control over the planet's phase state (at \SI{1.054}{\AU}). Cases more oxidising than Iron-W\"ustite always solidify, while more reducing cases are then subject to the planet's C/H ratio. An oxidation state comparable to modern Earth ($\sim$ IW+4) falls within the solidifying regime for all cases of $[H]$ explored in this work; geological evidence suggests that Earth's mantle has been oxidised since at least \mbox{3.9 Ga} \cite{rollinson_redox_2017, nicklas_redox_2018}. 
\par 

The cases at a Venus-like orbital separation of $\sim 0.737 \text{ AU}$ (not plotted in Figure \ref{fig:hydrogen_water}) vary between fully molten and mostly solidified. At this distance the models typically enter into a protracted mushy state following rapid initial solidification, which compares well with previous studies (e.g. \citeA{lebrun_thermal_2013}). Early Venus probably produced at least 2 oceans of \ce{H2O} from captured nebular \ce{H2}, placing $[H] \ge 2 \text{ oceans}$ and ruling out our $[H]=1\text{ oceans}$ case \cite{salvador_magma_2023, lammer_venus_2020}. Assuming that H capture is a viable avenue of water delivery to the young planet, at least 8 oceans of \ce{H2O} are expected to have been produced if Venus formed before protoplanetary disk dispersal \cite{salvador_magma_2023,williams_neon_2019,2020MNRAS.496.3755K}. However, geochemical evidence, such as the present-day D/H ratio of ocean water, suggest substantial uncertainty in these estimates \cite{broadley_origin_2022}. This leaves our $[H]=10\text{ oceans}$ case, which requires highly oxidising conditions ($\ge \text{IW} + 5$) and sufficiently low metallicities ($\text{C/H} > 0.5$) in order to avoid large amounts of remnant melt. This case (number 587) terminates within a pseudo-runaway greenhouse regime (OLR = \SI{364}{\WPMS}) with a thick ($P_s = \SI{2472}{\bar}$) \ce{CO2} dominated atmosphere which contains large amounts of steam (\ce{H2O} mole fraction of 31\%). The high surface temperature ($T_s = \SI{1740}{\kelvin}$) could cause most of the \ce{H2O} to escape during continued evolution, allowing the planet to desiccate and thereby replicate modern Venus-like conditions \cite{salvador_magma_2023}.

\subsection{Shallow magma oceans}
\label{sec:results_shallow} 
Since the models do not evolve past the point of solidification, surface temperatures do not cool below the surface solidus temperature. Across fully solidified cases, final net top of atmosphere fluxes range between approximate radiative equilibrium (\SI{3}{\WPMS}) and rapid energy loss (\SI{259}{\WPMS}). Cooling could allow for rainout of some volatile components, reducing the optical depth of the atmosphere.
\par
\begin{figure}[ht]
    \centering
    \includegraphics[width=\textwidth]{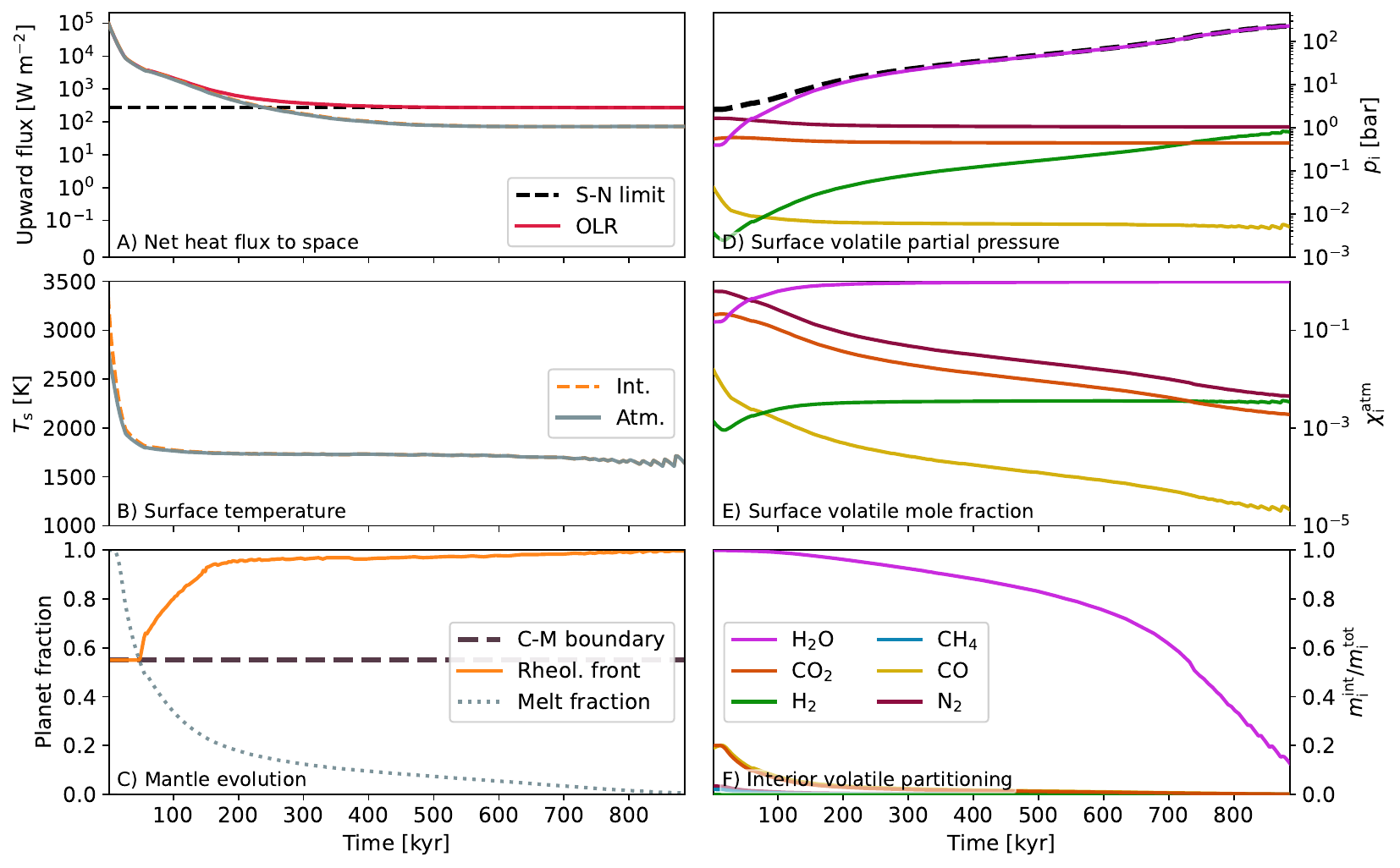}
    \caption{
    Time evolution of Case 714 ($a=1.054 \text{ AU}$, $\text{fO}_2 = \text{IW}+5$, $\text{C/H} = 0.01$, $[H] = 1.0$). This corresponds the upper-left corner of the left-column panels of Figure \ref{fig:hydrogen_water}, deep within the solidifying region of the grid. All panels share the same x-axis. The top left panel plots the interior (orange line) and atmospheric (grey line) net heat fluxes alongside the OLR. ``C-M boundary'' plots the location of the core-mantle boundary ($0.55 R_{\text{p}}$) as a baseline for the rheological front. The melt fraction is expressed as a fraction of the mantle -- panel c does not imply that the melt fraction of the core is evolving with time in these simulations.
    }
    \label{fig:case714}
\end{figure}

 
The initial gas chemical and dissolution equilibrium can yield relatively transparent atmospheres, enabling large net top of atmosphere fluxes and rapid cooling. These atmospheres contain little \ce{H2O} as it is dissolved in the magma, and in some cases, such as case 714 below, are \ce{N2} dominated. Self-regulation then occurs, with \ce{H2O} (and other volatiles) being outgassed due to bottom-up crystallisation of the mantle, which increases the radiative opacity of the atmosphere. This means that the majority of the interior typically crystallises rapidly until the optical depth of the atmosphere becomes large enough to stall -- or prevent, in radiative equilibrium cases -- solidification. One example of this behaviour is Case 714; Figure \ref{fig:case714} plots the evolution of the planet in this case.
Solidification starts from the bottom, with crystals forming at the core-mantle boundary and then at successively shallower layers as the magma ocean cools. The rheological front lifts off the core-mantle boundary at 48.9 kyr and exceeds $0.95 R_{\text{p}}$ by 185.0 kyr (Figure \ref{fig:case714}c). Crystallisation first occurs at the top-most layer after 26.7 kyr of evolution, so the magma ocean spends very little time in a completely liquid state, instead spending a significant period of the time with a shallow mixed-phase magma ocean.
\par 
Outgassing of \ce{H2O} is sufficient to shift the atmosphere from an \ce{N2} dominated composition into a \ce{H2O} dominated one (panel e). This also leads to increased abundances of \ce{H2} in the atmosphere through equilibrium chemistry. The radiative opacity of the atmosphere keeps the planet in a runaway regime ($T_{\text{s}} \sim 1700 \text{ K}$, panels b), with the OLR approximately equal to the Simpson-Nakajima limit (panel a) until solidification. The timescales of \ce{CO2} and \ce{CO} outgassing are shorter compared to \ce{H2O} due to their lower solubility in the mantle. Case 714 would continue to cool post-solidification, since the simulations terminates with a net positive top of atmosphere flux of \SI{72.6}{\WPMS} (Figure \ref{fig:case714}a). This case could remain within the runaway greenhouse regime for some period of time until the surface cools to a sufficiently low temperature (Figure \ref{fig:runaway}).
\par 
During the final stages of solidification ($\Phi < 1\%$) the shallow magma ocean near the surface exhibits non-linear variability on a short timescale of $\approx \SI{18}{\kilo\year}$, causing the surface temperature oscillate between 1640 and 1700 K (panel b) and outgassing to repeatedly pause and resume (panel f) due to the changing melt fraction. As the average melt fraction of the upper layers decreases, convection in the magma ocean becomes sluggish leading to inefficient upward energy transport. This traps heat energy below the surface. The local temperature increases as energy accumulates, eventually re-melting the layer and allowing convection to resume, cooling then continues once more. Volatiles are outgassed during the crystallising part of this cycle. This behavior is common to cases that fully solidify among our grid of 1029 models.

\subsection{Convective instability in the atmosphere}
\label{sec:results_convection}
Like most previous works, our atmosphere model uses a prescribed temperature structure where convection is assumed to be the primary energy transport process in the atmosphere \cite{lichtenberg_vertically_2021, hamano_lifetime_2015, kopparapu_habitable_2013}. However, in the case of pure steam atmospheres in equilibrium with incident stellar radiation, \citeA{selsis_cool_2023} showed that convection typically shuts down in the lower portions of thick atmospheres, because there is insufficient stellar flux penetrating to the surface to drive the atmosphere superadiabatic. These atmospheres develop a nearly isothermal layer near the planet's surface, similar to isothermal layers that appear in the analytic grey gas solution \cite{Guillot2010, robinson_analytic_2012, pierrehumbert_book_2010}. The results are subject to uncertainties in water vapor opacities at high pressure and temperature, but nonetheless require that the shutdown of convection be considered as a possibility.  When shutdown of convection in the lower atmosphere occurs, the surface temperature becomes lower than it would be on an adiabat. \citeA{selsis_cool_2023} did not consider the case of cooling of a primordial magma ocean, but in the context of a time-dependent calculation, convective shutdown would manifest itself as the lower atmosphere continuing to cool to temperatures below the adiabat, even after the top of atmosphere energy budget comes into balance.  This could lead to earlier solidification of a magma ocean than the adiabatic calculations predict, or could lead to solidification of magma oceans predicted to be permanent on the basis of a fully convective atmosphere.  In addition to considering the transient case where an initial magma ocean can provide sufficient flux to maintain convection, the configuration we treat differs from \citeA{selsis_cool_2023} in that we incorporate additional opacity sources that can block water vapor window regions and help maintain convection. 
\par
\begin{figure}[ht]
    \centering
    \includegraphics[width=\textwidth]{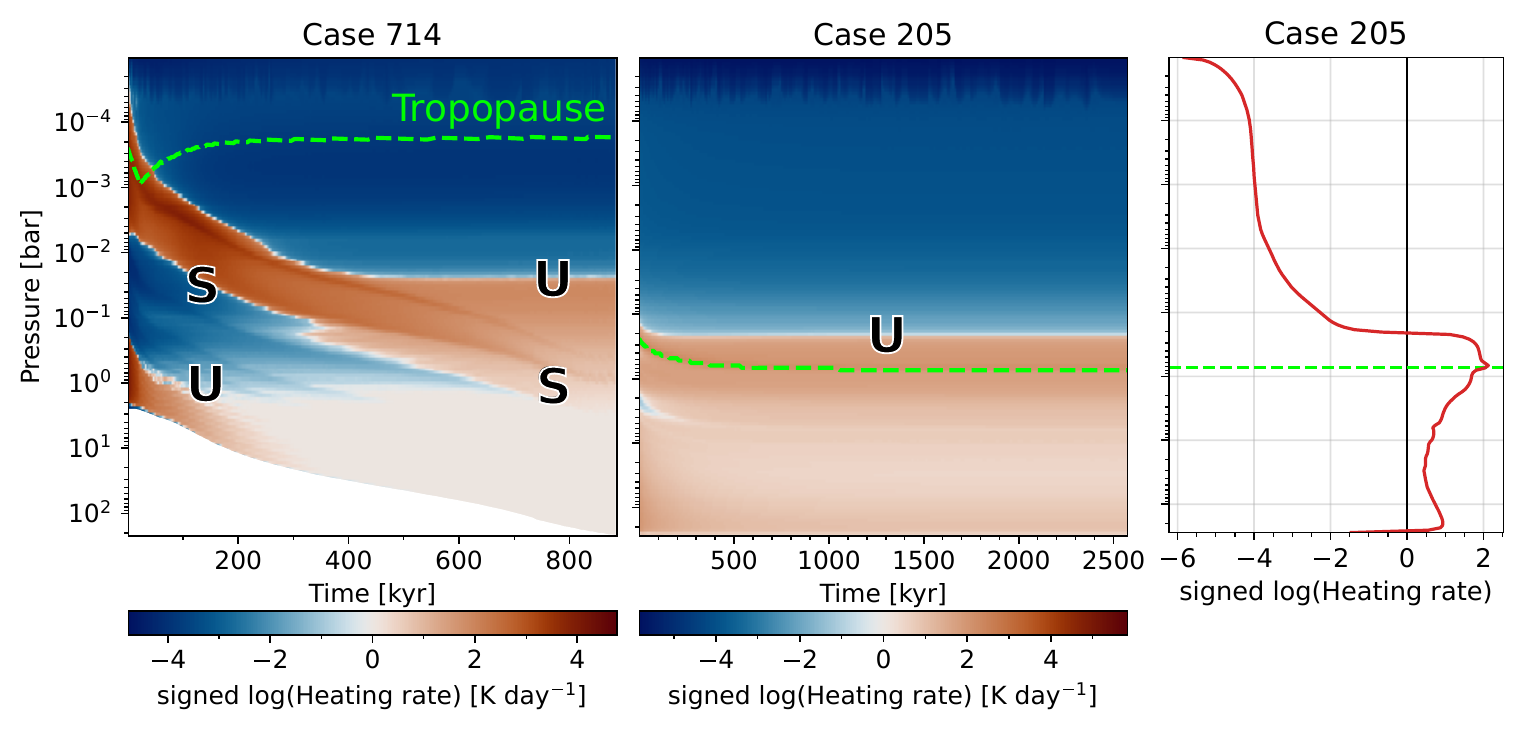}
    \caption{Radiative heating rate versus pressure and time. Case 714 (left) fully solidifies, and was discussed in Section \ref{sec:results_shallow}. Case 205 (centre) reaches global radiative equilibrium while maintaining a significant melt fraction of 97.6\% under a \ce{CO}- and \ce{H2}-dominated atmosphere. The dashed green lines indicate the tropopause. Regions of relative radiative heating and cooling are labelled `U' and `S', corresponding to whether or not they are unstable or stable to convection. The rightmost panel plots the radiative heating profile at the final timestep of Case 205.}
    \label{fig:convection}
\end{figure}

Ultimately, the possibility of convective shutdown needs to be addressed in a radiative convective model that allows for radiative layers to appear, by means of time-stepping the temperature profile or using a solver that finds atmospheric equilibria without assuming an adiabat. We will consider this point in a future study. As a preliminary indication of convective shutdown, we can test whether the adiabatic profile can be maintained by convection using a variant of the test employed in \cite{selsis_cool_2023}, which involves looking at vertical profiles of radiative heating (or equivalently net radiative flux) for an assumed adiabatic profile. Specifically, we can test whether or not our computed atmospheres would be unstable to convection by analysing the radiative heating rates of each layer ($H = (g/c_p)dF/dP$).  Radiative heating at deeper levels and cooling at adjacent upper levels would act to increase the lapse rate and promote convective instability, while the inverse configuration would act to decrease the lapse rate and lead to a convectively stable region.  Figure \ref{fig:convection} plots per-level radiative heating rates versus pressure, over time, for two of the cases explored in this work. The ``Tropopause'' marked in the figure indicates the boundary of the idealized isothermal stratosphere assumed in these calculations, whereas actual radiative equilibrium would in most cases yield non-isothermal stratospheres and shift the tropopause level.  The left panel indicates that Case 714 initially maintains strong convection from the surface due to the large surface temperature, but a radiative layer may be present at lower pressures ($p$ < 0.1 bar).  At solidification, Case 714 may not maintain convection from the surface due a small amount of radiative heating near 0.5 bar, although greater heating at lower pressures (0.01 bar) could trigger convection aloft. The centre panel indicates that Case 205 ($a=\SI{0.316}{\AU}$, $f\ce{O2}=\text{IW}-1$, $C/H=0.83$, $[H]=5$) remains convectively unstable near the surface and above the nominal tropopause throughout the simulation, including when it reaches global radiative equilibrium (rightmost panel). This suggests that the presence of additional opacity sources (in this case, primarily \SI{205}{\bar} of \ce{CO} and \SI{57} of \ce{H2}) may allow a permanent magma ocean to exist, though the suggestion must eventually be tested with more general radiative-convective equilibrium calculations. The right panel shows the radiative heating rate profile for the final timestep of Case 205 (middle panel): we can see that while convection is maintained near the surface and above the tropopause, the layer between roughly 10 bars and 1 bar tends to be stabilized by the heating profile. The resulting stable layer will be eroded by convection below and above, but whether the remaining stable sub-adiabatic lapse rate layer is sufficient to cause solidification must await tests with a full radiative-convective model. 

\section{Discussion}
\label{sec:discussion}

\subsection{Implications}
\label{sec:discuss_implications}
From these trends, we highlight a hierarchy of parameters to which the magma ocean evolution is sensitive. Orbital separation is the most important factor, followed by total hydrogen inventory, mantle oxygen fugacity, and finally the C/H ratio. It is possible that other factors not explored in this work would fit within this hierarchy:
\begin{itemize}
    \item Surface gravity. For hydrostatically supported atmospheres, surface pressure is proportional to surface gravity; so denser planets may have thicker atmospheres which are maintained against escape. 
    \item Cloud parameterisation. Formation of water clouds would reflect stellar flux and also induce additional greenhouse forcing, modulating the rate at which energy is escapes the planet.
    \item Stellar spectral class and age. Both of these factors can lead to variations in stellar bolometric emission, which directly impacts the planet's energy loss. Similarly, spectroscopic differences between stellar classes could impact cooling depending on the composition of the planet's atmosphere.
    \item Core size. This is directly tied to the volume of the mantle, and therefore the dissolved volatile inventory, surface pressure, and corresponding optical depth.
\end{itemize}
There is therefore potential for a very wide range of atmospheric compositions on lava planets, beyond even that exhibited by our modelling. We find that atmospheres on molten planets typically have a more diverse composition than on solidified planets (which are typically composed of \ce{H2} and/or \ce{H2O} at the end of our simulations). Additional post-solidification processes such as weathering may complicate this picture, but this remains applicable for young planets, such as TOI 1807 b and HD 63433 d \cite{Hedges_2021,capistrant_hd63433d_2024}. Telescope observations which constrain atmospheric composition of short-period planets could therefore be used to probe the interior properties and solidification regime of exoplanets. Recent observations by \citeA{hu_cancri_2024} indicate a carbon-rich \ce{CO2}-\ce{CO} atmosphere on the ultra-short period planet 55 Cancri e, which corresponds to a subset of our parameter space. Not only does this observation provide the first tentative evidence of a volatile-rich secondary atmosphere on an ultra-short period planet, it shows that secondary atmospheres can be maintained for Gyr periods despite the potential for ongoing escape processes. However, understanding compositional fractionation in the escape of H-rich atmospheres is likely key to resolving the link between a planet's magma ocean stage and its current conditions \cite{wordsworth_review_2022}.

\par 
HD 63433 d is a young Earth-sized exoplanet discovered in transit using TESS \cite{capistrant_hd63433d_2024}. Models and HST observations indicate that its outer neighbour (HD 63433 b) has lost its primordial H/He envelope \cite{zhang_escape_2022}. If the more irradiated planet (HD 63433 d) has also lost its envelope then the composition of an overlying secondary atmosphere is likely to be significantly influenced by volcanism or volatile exchange with a permanent magma ocean. The equilibrium temperature of this planet ($\sim \SI{1100}{\kelvin}$) is larger than that of the most irradiated cases ($a = \SI{1.054}{\AU} \implies T_{\text{eqm}}\sim\SI{885}{\kelvin}$, Figure \ref{fig:molten_pies}) modelled in this study, which indicates that it may maintain a permanent magma ocean. HD 63433 d can potentially provide an analogue for early Earth following the Moon-forming impact \cite{tonks_magma_1993, canup_moon_2001, warren_moon_1985}. The young age of this planet brings opportunities for direct inferences of planetary properties by comparison of observations with numerical models like PROTEUS. Future studies should consider the evolution and current state of this planet in greater detail.
\par 
With solidification comes outgassing of water (among other volatiles), shown in the bottom row of Figure \ref{fig:hydrogen_water} in which solidified cases have steamy atmospheres. Across solidified cases, the \ce{H2O} mole fraction varies smoothly with the oxygen fugacity and C/H ratio (bottom left panel) since atmospheric composition is no longer controlled by large amounts of melt. This behaviour is also present at higher $[H]$, but only in the absence of any significant amount of silicate melt since the \ce{H2O} will otherwise favourably dissolve into the interior of the planet. This makes the mole fraction of \ce{H2O} a useful tracer for a recently solidified surface, despite not being the only driver behind the phase state. This will be less applicable as the planet continues to evolve and the hydrogen escapes to space.

\par 
Gases beyond \ce{H2O} contribute to the opacity of these atmospheres, thereby impacting their cooling times or relaxation to radiative equilibrium. However, even for a pure-steam case we find that the post-runaway transition occurs at higher temperatures when modelled with the most up-to-date linelists, in contrast to models using HITRAN-derived opacities. Future work should explore the impacts of additional sources of atmospheric opacity. For example, \ce{CO} and \ce{H2}-\ce{H2O} collisional continua at high pressure and temperature \cite{karman_hitran_2019}.

\par 
It has been previously suggested that the mass, radii, and spectroscopic observations of sub-Neptune planets can be explained by the presence of magma oceans underneath thick \ce{H2} dominated atmospheres \cite{shorttle_magma_2024}. \ce{H2} dominated atmospheres are common on permanent magma ocean planets arising from our grid, so it may be possible that these sub-Neptunes have maintained molten surfaces since their formation with energy loss limited by atmospheric blanketing and sufficient instellations, confirming previous models with simpler atmospheric approaches \cite{2018ApJ...869..163V}. For example, case 608 has a reducing mantle and large hydrogen inventory ($a=\SI{1.054}{\AU}$, $f\ce{O2}=\text{IW}-5$, $C/H=2$, $[H]=10$) which allows it to reach global radiative equilibrium with an \ce{H2}-dominated atmosphere ($p_{\ce{H2}} = \SI{1038}{\bar}$) containing significant amounts of \ce{CO} (\SI{329}{\bar}) and \ce{CH4} (\SI{103}{\bar}). It is likely that such an atmosphere -- particularly one buffered by volatiles dissolved in the mantle -- could resist escape if formed around a sufficiently massive interior \cite{misener_escape_2021}. In such a case, the redox evolution of the interior would dictate the post-H-escape composition of its secondary atmosphere \cite{2021ApJ...914L...4L,2024arXiv240504057L}. 

\par 
It is unreasonable to assume that all planets (both within and external to the Solar System) have a similar or Earth-like mineralogy, redox state, and volatile endowment \cite{guimond_mineralogical_2023, putirka_polluted_2021, 2021SSRv..217...22G, 2024arXiv240504057L}. Our simulations show that -- at a fixed planet mass, instellation, and radius -- the evolution of a magma ocean planet is strongly dependent on the properties of its semi-molten interior \cite{meier_magma_2023}. This is directly reflected in the composition of the overlying atmosphere, both before and after the magma ocean solidifies (if it does, depending also on the properties of the mantle). It is therefore also unreasonable to assume that a given planet (mass, instellation, etc.) will be solidified or molten without intimate knowledge of these other parameters. However, polluted white dwarfs provide an opportunity to probe the mineralogy and elemental compositions of exoplanets and can therefore provide insight into which evolutionary pathways are compatible with contemporary observations of exoplanets \cite{xu_dwarf_2021,doyle_extrasolar_2019}. Observations of multiple exoplanets orbiting the same host star may also be helpful in lifting this degeneracy, since they will have closely related compositions but different instellations; characterising their present states and comparing with evolutionary models (such as PROTEUS) may thereby constrain their interior properties. Similarly, these generalised planetary evolution models can also be applied to planets within the solar system, where direct and in-situ measurements may provide further constraints on the physics \cite{2023SSRv..219...51S, 2024SpScT...4...75G}.
\par 
The results presented in Section \ref{sec:results_convection} indicate that these atmospheres on lava worlds may present radiative layers despite our assumption that they are fully convective below the stratosphere. This follows from the predictions of \citeA{selsis_cool_2023}. However, it is also shown that some planets which maintain permanent magma oceans (e.g. case 205 - Figure \ref{fig:convection}) are still able to remain convectively unstable in near-surface layers. Case 714, which outgasses a large steam atmosphere while solidifying, becomes weakly stable to convection near the surface under the thick steam atmosphere outgassed from the magma ocean. Calculation of an energy-conserving atmospheric temperature structure (and resultant energy fluxes) would also provide more accurate values for energy loss to space, potentially changing the solidification times calculated in this work \cite{2017JGRE..122.1458S}. Although JANUS prescribes a temperature profile for the atmosphere in a similar manner to previous works \cite{hamano_lifetime_2015, lebrun_thermal_2013, lichtenberg_vertically_2021, nikolaou_duration_2019}, it is possible that cases in our grid which do not solidify would do so in reality. Whether these atmospheres generally stay convective is seemingly not trivial to answer, so further investigation using a more comprehensive atmosphere model is merited. This should be explored for a range of atmospheric compositions generated from interior outgassing, ideally with the inclusion of compositional criteria for convective inhibition \cite{innes_runaway_2023}.

\subsection{Limitations}
\label{sec:discuss_limitations}

Our simulations show that cases at $a \le 0.527 \text{ AU}$ are completely dry at termination due to the combinations of high surface temperatures and comparatively low surface pressures from volatile dissolution. This means that volatile condensation does not occur, which precludes cloud formation but leaves open the possibility of photochemical haze production \cite{bergin_haze_2023,maillard_haze_2023}. Similarly, \citeA{lebrun_thermal_2013} found that rainout is unlikely to occur at $a < 0.66 \text{ AU}$ for planets orbiting the Sun. Shallow regions of moist \ce{H2O} convection occur for a subset of cases at \SI{0.737}{\AU}. At $a \ge 1.054 \text{ AU}$, moist convection and/or equilibrium saturation occurs in the upper atmosphere for almost all cases. This corresponds to solidified planets (panel k of Figure \ref{fig:ecdf_many}) where the mole fraction \ce{H2O} of water is large. Optically thick water clouds would reflect stellar radiation and absorb upwelling radiation, so it is unclear whether they would prolong or foreshorten magma ocean solidification \cite{pierrehumbert_book_2010,2017JGRE..122.1458S}. Rainout of \ce{H2O} could allow sufficient infrared transmission such that these planets escape the runaway greenhouse regime after some time. Future work should explore the role of aerosols in magma ocean evolution.
\par 

Atmospheric escape would act to remove volatiles from the system. However, due to the prescriptive nature of our atmosphere model, calculations of escape rate would offer little investigative value and are thus neglected. The equilibrium chemistry calculated for Case 714 (Figure \ref{fig:case714}) yields only minor abundances of low molecular weight species (\ce{H2}, \ce{CO}), so it is possible that this atmosphere would be maintained against significant escape for the duration of this simulation. However, more complete calculations of the chemistry would be required to determine the speciation of elements in the regions of the upper atmosphere from which escape is occurring  \cite{tsai_chemistry_2021,2023MNRAS.523.5681N}. Together, chemistry and escape could influence outgassing rates -- subject to sufficient mixing -- through transport of volatiles from the surface into the aloft regions, which is then compensated by further outgassing \cite{wordsworth_redox_2018}. Chemistry and escape can also act to distil the planet's total volatile content through the loss of low molecular weight species, thereby reducing the opacity and increasing radiative emission to space \cite{wordsworth_redox_2018, 2024ApJ...967..139C}. Together, chemistry and escape are expected to shorten magma ocean lifetimes compared to the calculations made in this work. Uncertainties in volatile accretion, however, could partially compensate for these processes \cite{zahnle_impact_2020, 2020A&A...643L...1V, 2022ApJ...938L...3L}. Future work should consider simulating escape processes self-consistently with magma ocean evolution, particularly in the context of highly irradiated rocky planets around M-type stars such as TRAPPIST-1b \cite{zieba_obs_2023}.
\par

We do not include tidal effects on the interior which could be significant source of heating for planets within tightly packed systems or on elliptical orbits \cite{bolmont_tidal_2013, seligman_tidal_2024, bower_linking_2019,hay_tides_2019}. Our interior model, which is able to resolve melt-crystal separation as the magma ocean evolves, shows that interior energy transport by convection is largely offset by other transport processes. In the case of tidal heating dissipated near the surface, this could act to extend magma ocean lifetimes and potentially expose a greater fraction of a planet's volatile inventory to escape.

\par 
In this work we use a mass-balance approach to simultaneously solve for chemical and dissolution equilibrium, but only for a small set of volatiles. Rock vapours (e.g. \ce{SiO}) are expected to be present in the atmospheres of magma ocean planets to various degrees \cite{2023ApJ...947...64W}, potentially impacting their radiation environment and inducing upper-atmosphere thermal inversions \cite{piette_rock_2023,zilinskas_rock_2023}. The latter two works solve for chemical and dissolution equilibrium as separate processes, obtaining the total gas budgets by adding the results of each calculation. \citeA{falco_rock_2024} self-consistently solved for chemical and dissolution equilibrium of rock vapours and volatiles, finding that it is necessary to solve for the whole system simultaneously. Both \citeA{falco_rock_2024} and \citeA{zilinskas_rock_2023} found that only small hydrogen budgets are required to prevent the formation of a thermal inversion in the atmosphere, since increased amounts of hydrogen promotes the abundance of \ce{H2O} and resultant infrared absorption. The transition between inverted and non-inverted temperature profiles is narrow. Using the empirical fit for the transition derived by \citeA{falco_rock_2024} we find that all cases explored in our grid fall deep within the non-inverted regime. It is therefore expected that the inclusion of rock vapours would have only a minor impact on our results. 

\par 
Finally, our simulations do not account for other erosion or delivery processes, such as those from giant impacts, which in the Solar System are though to have contributed to shaping the final volatile budget of the terrestrial planets \cite{2018SSRv..214...34S}. However, estimates of volatile losses by impact erosion differ by many orders of magnitude \cite{2020ApJ...901L..31K,2022MNRAS.513.1680D,2024PSJ.....5...28L} and it is further unclear if late-accreting planetesimals would instead be a source of volatiles themselves \cite{2022ApJ...938L...3L}. Hence, in this work we choose to restrict the volatile content to be fixed at the onset of our simulations, but future work shall consider a time-varying volatile accretion history during early thermal planetary evolution.

\section{Conclusions}
\label{sec:conclusions}
We have developed a coupled numerical model to simulate the rapid cooling stage of molten planets at arbitrary oxidation state, resulting in cases which either fully solidify or reach global radiative equilibrium while maintaining some amount of melt. Our conclusions are as follows.
\begin{enumerate}
    \item The atmospheres which overlie magma oceans can have diverse compositions ranging from \ce{H2}- to \ce{CO2}-dominated, while recently solidified planets typically have atmospheres composed of \ce{H2} and/or \ce{H2O} (by mole fraction).
    
    \item Thermal blanketing by radiatively opaque atmospheres means that magma oceans spend most of their lifetime in a semi-molten state, with significant energy transport generated by crystallisation and gravitational settling. Collisional absorption by \ce{H2} is key to moderating magma ocean cooling, extending their lifetimes by 10s to 100s of Myr, even around Sun-like stars.

    \item Whether or not a magma ocean solidifies depends not only on its instellation, but also on other properties such as the bulk C/H ratio, mantle oxygen fugacity, and bulk hydrogen inventory. It is thus critically important to consider a wide range of initial conditions when using observations of a planet to constrain its historical evolution.

    \item Radiative heating rates within these model atmospheres indicate that they maintain strong atmospheric convection while very hot and molten. However, it is possible that this convection shuts down at the surface at later points in their evolution. Future work should explore this in detail.
\end{enumerate}
\par 
Our numerical framework enables future modelling to incorporate a range of additional physics self-consistently, such as: radiative-convective energy balance and chemical kinetics in the atmosphere, escape processes, tidal heating, stellar evolution, impacts and volatile delivery, and geochemical cycling. Inclusion of a number of these properties will enable more complete exploration of the geophysical history of exoplanets in diverse environments. These avenues will be explored in future work. We particularly emphasise the importance of realistic atmospheric temperature structures and real-gas radiative transfer.

\section{Open Research}
The source code for PROTEUS (\citeA{nicholls_proteus_2024} --- Zenodo DOI: 10.5281/zenodo.12190523, \url{https://github.com/FormingWorlds/PROTEUS}) and JANUS (\citeA{nicholls_janus_2024} --- Zenodo DOI: 10.5281/zenodo.12190459, \url{https://github.com/FormingWorlds/JANUS}) are available under the Apache 2.0 license. The source code for SPIDER \cite{bower_retention_2022} is available on GitHub (\url{https://github.com/djbower/spider}) under the GPL 3.0 license. The source code for SOCRATES with additional tools for handling opacities (\citeA{nicholls_socrates_2024} --- Zenodo DOI: 10.5281/zenodo.12190852, \url{https://github.com/nichollsh/SOCRATES}) is available under the BSD 3-clause license. MORS \cite{johnstone_active_2021} is available on GitHub (\url{https://github.com/FormingWorlds/MORS}) under the MIT license. Opacity data \cite{grimm_database_2021} are available online on the University of Geneva DACE website. 
\par 
Case-by-case data and plots, additional analysis software, and the figures used in this article are available under the Apache 2.0 license via the Open Science Framework (\citeA{osf_repository} --- DOI: 10.17605/OSF.IO/4EQHN). 
\par 
Analysis was performed with Python, Numpy \cite{harris_numpy_2020}, Scipy \cite{pauli_scipy_2020}, and Matplotlib \cite{hunter_matplotlib_2007}. Data were represented using the Scientific Colour Maps \cite{crameri_colours_2023}.


\acknowledgments
We thank our two anonymous reviewers for their valuable feedback which significantly improved this paper.
\par 

CRediT author statement. \textbf{Harrison Nicholls}: Conceptualization, Methodology, Software, Validation, Investigation, Writing - Original Draft, Writing - Review \& Editing, Visualization, Data Curation. \textbf{Tim Lichtenberg}: Conceptualization, Writing - Review \& Editing. \textbf{Dan Bower}: Software, Writing - Review \& Editing. \textbf{Raymond Pierrehumbert}: Conceptualization, Resources, Writing - Review \& Editing, Supervision.
\par 

H.N. was supported by the Clarendon Fund and the MT Scholarship Trust. T.L. was supported by the Branco Weiss Foundation, the Alfred P. Sloan Foundation (AEThER project, G202114194), and NASA's Nexus for Exoplanet System Science research coordination network (Alien Earths project, 80NSSC21K0593).


\bibliography{main.bib}

\end{document}